\documentclass[12pt,letterpaper]{article}
\pdfoutput=1

\usepackage[utf8]{inputenc}
\usepackage[UKenglish]{babel}
\usepackage[dvipsnames]{xcolor}
\usepackage{amsmath,amssymb,mathtools}
\usepackage{braket}

\usepackage[left=2cm,top=1cm,right=3cm,nohead]{geometry}
\usepackage{setspace,footmisc}
\usepackage[toc,page]{appendix}
\renewcommand{\baselinestretch}{1.2}
\numberwithin{equation}{section}

\usepackage{array,longtable,booktabs}
\usepackage{graphicx,float}
\usepackage[bf]{caption}
\usepackage{tikz}

\usepackage{cite}
\usepackage[
    colorlinks=true,
    linkcolor=Blue,
    citecolor=Blue,
    filecolor=Blue,
    urlcolor=Blue,
    linktoc=page,
    pdfstartview=FitV,
    bookmarksopen=true
]{hyperref}

\begin{document}

\pagestyle{empty}
\begin{center}
{\bf\LARGE 4D $\mathcal{N} = 4$ String Islands and S-duality\\ [3mm]}

\large{H\'ector Parra De Freitas$^{1,4}$ and Houri-Christina Tarazi$^{2,3,4}$
\\[2mm]}

{\small ${}^1$ Mathematical Sciences \& STAG Research Centre, University of Southampton\\ [-1mm]}
{\small\textit{Southampton, SO17, UK}\\[0.2cm]}
{\small ${}^2$ Kavli Institute for Cosmological Physics, University of Chicago\\ [-1mm]}
{\small\textit{Chicago, IL 60637, USA}\\[0.2cm]}
{\small ${}^3$ Leinweber Institute for Theoretical Physics, University of Chicago\\ [-1mm]}
{\small\textit{Chicago, IL 60637, USA}\\[0.2cm]}
{\small ${}^4$ Jefferson Physical Laboratory, Harvard University\\ [-1mm]}
{\small\textit{Cambridge, MA 02138, USA}\\[0.2cm]}

\vspace{0.3in}

\small{\bf Abstract} \\[3mm]
\end{center}
String islands with four-dimensional $\mathcal N=4$ supersymmetry have no compactification moduli; their only
continuous parameter is the complex axiodilaton. We construct a family of heterotic and type II islands as
asymmetric orbifolds using symmetries of the Leech and $E_8$ lattices, respectively, and investigate their
S-duality relations. Assuming that the gravitational thresholds retain the generic form for CHL models and
that perturbative electric charges determine the full electromagnetic lattice, we obtain candidate duality
orbits and test them by explicit lattice computations. The electric charge lattices close under the predicted
Atkin--Lehner operations, providing evidence for an extension of the duality patterns of CHL models found by
Persson and Volpato to intrinsically four-dimensional models. In particular, we identify a heterotic S-dual of
the type II 4D island of Dabholkar and Harvey. For each symmetry, the charge lattices form a complete genus,
which organizes the corresponding family of islands and their proposed dualities.

\newpage

\setcounter{page}{1}
\pagestyle{plain}
\renewcommand{\thefootnote}{\arabic{footnote}}
\setcounter{footnote}{0}

\tableofcontents

\section{Introduction}

String islands are perturbative string compactifications with only the dilaton as a perturbative scalar
modulus~\cite{Dabholkar:1998kv}. Dabholkar and Harvey introduced these theories in connection with the problem
of vacuum degeneracy and the phenomenological and cosmological difficulties associated with massless moduli.
The tendency of the dilaton to run towards zero coupling in attempts to break supersymmetry gives a further
motivation for studying vacua with few moduli. By removing the compactification moduli, string islands address
part of the moduli-stabilization problem and provide realizations of pure supergravity in string theory. They
also raise the question of whether strong coupling necessarily leads to a higher-dimensional
description.\footnote{The broader goal of removing or stabilizing moduli has motivated several complementary
approaches. Recent developments include fully stabilized $\mathcal N=1$ Minkowski
solutions~\cite{Becker:2024ayh,Rajaguru:2024emw,Chen:2025rkb}, nonsupersymmetric heterotic constructions
without neutral compactification moduli~\cite{Baykara:2024tjr,Angelantonj:2024jtu,Basaad:2024lno}, and
isolated nonsupersymmetric critical points~\cite{Chen:2025rkb,Baykara:2026jzs}.}

String islands have been constructed in both six and four dimensions. A 6D $\mathcal N=(1,1)$ type II example
was obtained as a $\mathbb Z_5$ asymmetric orbifold in~\cite{Dabholkar:1998kv}. The predictions
of~\cite{Fraiman:2022aik} subsequently led to the classification and construction of all 6D $\mathcal N=(1,1)$
type II islands in~\cite{Baykara:2025lhl} and the construction of all heterotic islands
in~\cite{Aldazabal:2025zht}. In four dimensions, the first examples include the heterotic model
of~\cite{Chaudhuri:1995fk}, obtained in the fermionic construction with no vector multiplets, and the type II
$\mathbb Z_9$ asymmetric orbifold of~\cite{Dabholkar:1998kv}. Further four-dimensional type II constructions
were studied in~\cite{Mizoguchi:2001cp}, and models with no vector multiplets were also obtained through
asymmetric Gepner constructions~\cite{Blumenhagen:2016rof}. A special class of heterotic islands in 4D,
forming an archipelago (see below), was constructed as well in~\cite{Aldazabal:2025zht}.

These constructions define the islands at weak string coupling, leaving open what becomes of them as the
dilaton moves towards strong coupling. In theories with sixteen supercharges, this limit often admits a
description in terms of another string theory or a higher-dimensional theory. Determining whether such a
description exists for string islands, and what form it takes, motivates our study of their dualities.

To identify the dual frames of these islands, it is useful to examine how strong-coupling limits are realized
in other four-dimensional theories with sixteen supercharges. Gregori et al.~\cite{Gregori:1997hi} exhibited
S-dualities relating perturbative heterotic and type II asymmetric orbifolds, and Persson and
Volpato~\cite{Persson:2015jka} organized a broad family of such relations for CHL models. These examples
suggest that the strong-coupling limits of four-dimensional islands may likewise admit perturbative heterotic
or type II descriptions. We propose that this duality pattern extends to string islands.

We consider asymmetric orbifolds defined by symmetries transported from the Leech lattice in the heterotic
case and from the $E_8$ lattice in the type II case.\footnote{Throughout our construction and S-duality analysis,
we restrict to symmetries that satisfy level matching before the shift is introduced. In the notation
of~\cite{Persson:2015jka}, this is the condition $\lambda=1$, where $\lambda$ is the order of the
level-matching anomaly, defined in~\eqref{eq:lambda-definition}. The models considered here therefore
form a subset of the possible island constructions.} The heterotic construction follows the approach
of~\cite{Aldazabal:2025zht}. These orbifolds extend the lattice-symmetry construction underlying the
perturbative frames studied in~\cite{Persson:2015jka} to symmetries that remove all compactification moduli.
The removal of these moduli also prevents us from applying their duality argument directly.

For the CHL models, Persson and Volpato use a type IIA orbifold frame on $K3\times T^2$, in which
four-dimensional S-duality becomes T-duality of the additional $T^2$. This frame retains torus moduli
and is therefore unavailable for islands. To test whether the CHL duality pattern nevertheless extends
to our constructions, we use an ansatz for the gravitational $R^2$ couplings to predict
which descriptions should be related, and test these predictions on electromagnetic charge lattices. The
latter can be determined perturbatively under the assumptions stated in
Section~\ref{sec:charge-lattice-duality}.

\subsection*{Overview of main results and their consequences}
\label{sec:summary-results}

Our results support a picture in which strong coupling relates four-dimensional string islands to other
perturbative string descriptions, despite the absence of compactification moduli. The proposed dualities
include a heterotic description of the order-nine type II island of~\cite{Dabholkar:1998kv}, and extend
throughout the family constructed here. In every case, the electromagnetic charge lattices transform as
predicted by the gravitational thresholds. This agreement also gives evidence for the thresholds themselves.
Their full $R^2$ couplings require nonperturbative methods that are not presently available for these theories.
We propose their form through an ansatz, whose duality transformations are then tested against independently
constructed charge lattices. The agreement thus provides a prediction for a future nonperturbative
calculation, without requiring a dual frame in which the coupling can already be computed.

Identifying the dual descriptions requires both ingredients. In the order-six example associated with the
Leech-lattice symmetry labelled $-6F$, a heterotic island is related to three type II descriptions, but these
four frames have only two distinct charge lattices. The gravitational thresholds distinguish the frames and
determine which one is reached by each Atkin--Lehner operation. Even Fricke invariance of a charge lattice
need not imply self-duality of the theory. For example, the three $12$-modular lattices in the $-12D$ archipelago are
preserved while Fricke duality exchanges the heterotic and type II descriptions. Their different threshold
frame shapes reveal this exchange, which the charge lattices alone cannot detect. 

There is a complementary
ambiguity when a single symmetry admits several charge lattices. The resulting family, which we call an \emph{archipelago}
following~\cite{Aldazabal:2025zht}, can be permuted by a duality that leaves the threshold unchanged. The
order-thirty example exhibits precisely this behavior. Together, the thresholds and charge lattices allow us
to follow the dualities down to individual members of an archipelago.

Taking these dualities seriously has a further consequence for the number of distinct islands. Inequivalent
shift vectors can give the same charge lattice, yet their multiplicities need not match between dual frames.
For example, the heterotic $9B$ construction has three such shifts, while its proposed type II dual has only one. If the
duality holds, these counts cannot both count distinct theories unless there are additional degeneracies
that our construction has missed. The most parsimonious explanation is that different shifts can describe
the same CFT. S-duality therefore directly constrains the apparent multiplicity of orbifold constructions,
and suggests the stronger conjecture that, for a fixed symmetry, all shifts giving isometric charge lattices
are equivalent. Such redundancy already occurs in pure shift orbifolds of lattice CFTs. The analysis
of~\cite{Israel:2025ouq} gives explicit equivalences between different shift presentations, even with
different cyclic orbifold orders, and applies them to shift orbifolds of the Leech lattice. A further
precedent is provided by the different cyclic gaugings of the Leech lattice theory that yield the same
Moonshine module~\cite{Carnahan:2017pnh}. We find evidence within
our own construction at order twenty-three, where shifts with the same charge lattice have identical unrefined
one-loop partition functions, before supersymmetric cancellation. This tests their full perturbative spectra. A proof of CFT equivalence would additionally have to identify their operator products.

Table~\ref{tab:intro-archipelagos} summarizes the families constructed here and their proposed duality
relations. We denote the heterotic and type II descriptions by $\mathrm{HET}_n$ and $\mathrm{II}_n$,
respectively, where $n$ is the symmetry order, including the Spin lift in the type II case. Primes distinguish
descriptions of the same type and order, following their order of appearance in
Table~\ref{tab:charge-lattice-classification}.

\begin{table}[htbp]
    \centering
    \small
    \renewcommand{\baselinestretch}{1}\selectfont
    \setlength{\tabcolsep}{4pt}
    \renewcommand{\arraystretch}{1.3}
    \begin{tabular}{@{}lc@{\hspace{1.5em}}lc@{}}
        \toprule
        S-duality orbits & Islands & S-duality orbits & Islands \\
        \midrule
        $\mathrm{II}_{4}$ & 1 & $\mathrm{HET}_{15},\ \mathrm{II}_{15}$ & 9 \\
        $\mathrm{HET}_{6},\ \mathrm{II}_{6},\ \mathrm{II}_{6}^{\prime},\ \mathrm{II}_{6}^{\prime\prime}$ & 1 & $\mathrm{HET}_{16}$ & 2 \\
        $\mathrm{II}_{8}$ & 4 & $\mathrm{HET}_{18},\ \mathrm{HET}_{18}^{\prime},\ \mathrm{HET}_{18}^{\prime\prime},\ \mathrm{II}_{18}$ & 3 \\
        $\mathrm{HET}_{8},\ \mathrm{II}_{8}^{\prime}$ & 1 & $\mathrm{HET}_{20},\ \mathrm{II}_{20}$ & 19 \\
        $\mathrm{HET}_{9},\ \mathrm{II}_{9}$ & 1 & $\mathrm{HET}_{20}^{\prime},\ \mathrm{II}_{20}^{\prime}$ & 11 \\
        $\mathrm{HET}_{12},\ \mathrm{II}_{12}$ & 3 & $\mathrm{HET}_{23}$ & 29 \\
        $\mathrm{HET}_{12}^{\prime},\ \mathrm{II}_{12}^{\prime}$ & 2 & $\mathrm{HET}_{24},\ \mathrm{II}_{24}$ & 35 \\
        $\mathrm{HET}_{12}^{\prime\prime},\ \mathrm{II}_{12}^{\prime\prime},\ \mathrm{II}_{12}^{\prime\prime\prime},\ \mathrm{II}_{12}^{\prime\prime\prime\prime}$ & 5 & $\mathrm{HET}_{24}^{\prime},\ \mathrm{II}_{24}^{\prime}$ & 29 \\
        $\mathrm{HET}_{12}^{\prime\prime\prime},\ \mathrm{II}_{12}^{\prime\prime\prime\prime\prime}$ & 3 & $\mathrm{HET}_{28}$ & 35 \\
        $\mathrm{HET}_{14},\ \mathrm{II}_{14}$ & 8 & $\mathrm{HET}_{30},\ \mathrm{HET}_{30}^{\prime},\ \mathrm{HET}_{30}^{\prime\prime},\ \mathrm{II}_{30}$ & 40 \\
        \bottomrule
    \end{tabular}
    \caption{The twenty proposed duality orbits of symmetry descriptions, arranged in two blocks.
    Each entry lists the descriptions in one orbit and the number of islands in each archipelago,
    assuming that distinct islands for a fixed symmetry are counted by inequivalent electromagnetic charge
    lattices.}
    \label{tab:intro-archipelagos}
\end{table}

The charge-lattice classification also suggests how genuinely distinct members of an archipelago might be
related. For each symmetry description, the lattices fill a complete genus and become isometric after circle
compactification. They can therefore be viewed, at the level of charges, as different decompactification
limits of a common three-dimensional theory. When an island is individually fixed by Fricke duality, its
charge lattice must have an $n$-modular positive-definite dual of
rank six. Our constructions yield 121 distinct lattices of this kind, including a complete genus of
twenty-nine $23$-modular lattices. Many appear to be new, with 119 having no matching entry in the Nebe--Sloane
catalogue records examined in Appendix~\ref{app:modular-catalogue}. The duality analysis thus produces a
collection of modular lattices whose mathematical interest extends beyond their role as string charges.

The resulting duality orbits also sharpen the question of which string descriptions are needed to describe
four-dimensional vacua with sixteen supercharges. Some of our heterotic islands have only one
infinite-distance limit up to duality, and hence only the perturbative heterotic asymptotic frame. If a
universal string frame exists for these vacua, it must therefore be heterotic. We explore this implication
in an upcoming paper~\cite{ParraDeFreitasTarazi:Upcoming}. For other islands, an open problem arises from cusps at non-square-free orders that lie
outside the Atkin--Lehner orbit of the known perturbative description. Already for the order-four type II
island, the interpretation of such a limit remains unknown. Our results identify and test a family of
strong-coupling descriptions while leaving these additional limits to be understood.

The paper is organized as follows. Sections~\ref{sec:heterotic-construction} and~\ref{sec:typeII-construction}
describe the heterotic and type II constructions and determine their perturbative electric charge lattices.
Section~\ref{sec:s-duality} reviews the CHL duality argument, formulates its extension to islands, and derives
the threshold and charge-lattice tests. Section~\ref{sec:explicit-results} presents the classification and
representative examples. We discuss the implications and open questions in
Section~\ref{sec:discussion-results}. The appendices collect lattice conventions, the shift-enumeration
method, explicit lattice data, and the comparison with the catalogue of modular lattices.

\noindent\textit{Note on related work.} Our work is independent of but overlaps with the recent work of Aldazabal et al.~\cite{Aldazabal:2026hix} in the construction of four-dimensional heterotic and type II islands. 

\section{Construction of heterotic islands}
\label{sec:heterotic-construction}

In this section we construct heterotic islands from Leech-lattice symmetries and determine their perturbative
electric charge lattices. We use the asymmetric-orbifold framework of~\cite{Narain:1986qm,Narain:1990mw}. The
construction requires a compatible Narain lattice and a shift that removes massless twisted-sector states. Our
conventions for lattices, discriminant forms and gluing are collected in
Appendix~\ref{app:lattice-definitions}.

We begin with the heterotic string compactified on $T^6$, so that the internal momenta take values in an even
self-dual Narain lattice $\Gamma_{22,6}$~\cite{Narain:1985jj,Narain:1986am}. To obtain an orbifold preserving
all sixteen supercharges, we choose an order-$n$ automorphism $\Theta$ that acts trivially on the right-moving
directions. For an island, its action must also remove all left-moving invariant directions, which would
otherwise give vector multiplets. We denote by
\begin{equation}
    I=\{P\in \Gamma_{22,6}\mid \Theta P=P\}
\end{equation}
the invariant lattice, and by
\begin{equation}
    N=I^\perp\subset \Gamma_{22,6}
\end{equation}
the normal (or coinvariant) lattice. For the rank-zero models considered here, $I$ has signature $(0,6)$,
while $N$ has signature $(22,0)$. Equivalently, all left-moving internal directions are rotated by $\Theta$.

The orbifold generator $g$ acts on the untwisted Hilbert space by
\begin{equation}
    g\ket{P_N,P_I}
    =
    e^{2\pi i P\cdot v}\ket{\Theta P_N,P_I},
\end{equation}
where the shift $v\in I\otimes \mathbb Q$ is taken along the invariant directions. The shift is chosen to
ensure modular invariance and to exclude massless states in the twisted sectors. These conditions depend on
the invariant lattice because the trace with insertion of $g$ receives momentum contributions from $I$, and a modular
transformation gives the $g$-twisted momenta in $I^*+v$. We therefore first determine the allowed symmetries
and their invariant lattices, then identify the admissible shifts and compute the resulting electric charge
lattices.

\subsection{Symmetries from the Leech lattice}

The coinvariant lattice provides a convenient way to specify the symmetry and its integral realization.
Consider an even Euclidean lattice $N$ of signature $(s,0)$, together with a primitive embedding
\begin{equation}
    N\hookrightarrow\Gamma_{22,6}.
\end{equation}
Let $\Theta\in O(N)$ be an automorphism acting trivially on the discriminant group
\begin{equation}
    A_N=N^*/N.
\end{equation}
Then $\Theta$ extends to an automorphism of $\Gamma_{22,6}$ by acting as the identity on the orthogonal
complement of $N$. Thus, a primitive embedding of $N$, together with such an automorphism, provides the
required Narain lattice symmetry. In the present work we specialize to $s=22$, with $\Theta$ rotating all
directions in $N$, as required for the rank-zero models.

We further require $N$ to be rootless. Otherwise, its roots would form orbits under $\Theta$, giving rise to
massless vector multiplets in the orbifold theory. We are therefore interested in even, rootless lattices of
rank 22 admitting a primitive embedding into $\Gamma_{22,6}$ and automorphisms that rotate all directions
while acting trivially on $A_N$.

A family of lattices with these properties can be obtained from symmetries of the Leech lattice
$\Lambda_{\mathrm{Leech}}$. Let $\Theta'\in O(\Lambda_{\mathrm{Leech}})$ have a rank-two invariant lattice,
and denote its invariant and normal lattices by
\begin{equation}
    I^{\mathrm{Leech}}=\Lambda_{\mathrm{Leech}}^{\Theta'},
    \qquad
    N=(I^{\mathrm{Leech}})^\perp\subset\Lambda_{\mathrm{Leech}}.
\end{equation}
Then $N$ is even, rootless and of rank 22. We will show that it admits a primitive embedding into
$\Gamma_{22,6}$, for which the restriction of $\Theta'$ to $N$ extends to a symmetry acting trivially on the
orthogonal complement.

Since the Leech lattice is even and self-dual, the discriminant forms of its primitive orthogonal sublattices satisfy
\begin{equation}
    q_N\simeq-q_{I^{\mathrm{Leech}}}.
\end{equation}
From standard lattice embedding theorems, any even positive-definite lattice of rank two admits a primitive
embedding into $E_8$. Choose such an embedding of $I^{\mathrm{Leech}}$, and let $I^{+}$ be its orthogonal
complement in $E_8$. We use the superscript $+$ for the positive-definite representative of the Narain
invariant lattice, so that $I^{+}=I(-1)$, while the superscript $\mathrm{Leech}$ distinguishes the fixed lattice in
the Leech lattice. The complement $I^{+}$ has rank six, and self-duality of $E_8$ gives
\begin{equation}
    q_{I^{\mathrm{Leech}}}\simeq-q_{I^{+}},
    \qquad\Longrightarrow\qquad
    q_N\simeq q_{I^{+}}.
\end{equation}
Thus $N$ and $I^{+}(-1)$, where $(-1)$ denotes reversal of the lattice pairing, have opposite discriminant
forms. Gluing them along their discriminant groups gives an even self-dual lattice of signature $(22,6)$,
hence a lattice isometric to $\Gamma_{22,6}$. This provides the desired primitive embedding of $N$, with
orthogonal complement $I^{+}(-1)$.

We organize the constructions by the isometry class of the invariant lattice $I$ and do not consider
possible degeneracies arising from inequivalent gluings at fixed $(N,I)$. Such degeneracies would give
the same possible charge lattices, since the construction in Section~\ref{ss:chargelattices} depends only
on $I$ and the shift $v$, through its kernel in $I$.

Finally, an automorphism of $N$ extends to either the Leech lattice or $\Gamma_{22,6}$, acting as the identity
on the respective orthogonal complement, precisely when it acts trivially on $A_N$. The extension conditions
are therefore the same in both cases. In particular, the restriction of $\Theta'$ to $N$ defines a Narain
lattice symmetry $\Theta$ of the required type.

\subsection{Shift vectors}
\label{ss:shift-vectors}

\paragraph{Level matching and the anomaly parameter.}
Before adding a shift, the chosen order-$n$ lift of $\Theta$ can have a level-matching obstruction. Write the
left-minus-right vacuum-energy mismatch in its twisted sector as~\cite{Persson:2015jka}
\begin{equation}\label{eq:lambda-definition}
 \Delta E_\Theta\equiv\frac{\mathcal E_\Theta}{n\lambda}
 \pmod{\tfrac1n\mathbb Z},\qquad
 \gcd(\mathcal E_\Theta,\lambda)=1.
\end{equation}
Thus $\lambda$ is the denominator of $n\Delta E_\Theta$ in lowest terms. The symmetry-dependent multiplier of
the $\Theta$-twisted torus trace under $T^n:\tau\mapsto\tau+n$ is
\begin{equation}
 \epsilon_\Theta=\exp(2\pi i n\Delta E_\Theta)
 =\exp\!\left(\frac{2\pi i\mathcal E_\Theta}{\lambda}\right).
\end{equation}
Its order is $\lambda$. In particular, $\lambda=1$ means that the unshifted symmetry satisfies level matching.
For $\lambda>1$, the phase obstructs gauging $\Theta$ alone. This phase measures the finite-symmetry anomaly
of $\Theta$ through worldsheet modular covariance.

Our restriction to level-matched symmetries is motivated by the CHL setting with the standard order-$n$ shift
and the electric--magnetic decomposition that we extrapolate to islands. In the CHL presentation, the shift
can be a null translation along a spectator circle with the same order as $\Theta$. For $\lambda>1$, an
order-$n$ shift must compensate the mismatch through its norm. Alternatively, Persson and Volpato use a null
shift of order $n\lambda$, giving the combined generator effective order $\widehat N=n\lambda$. Their general
electromagnetic lattice then requires additional data and need not have the simple purely electric--purely
magnetic splitting used here (Section~2.3 and Appendix~A.4 of~\cite{Persson:2015jka}). In our islands the
invariant lattice is definite, so the shifts lie in that lattice and obey the norm condition below. The
standard feature retained is the order $n$ and the absence of a mismatch to compensate.

For each choice of symmetry $\Theta$, we must determine the allowed shift vectors $v\in I\otimes\mathbb Q$. We
first impose modular invariance and the absence of massless states in the twisted sectors, following the
analysis of \cite{Aldazabal:2025zht}. We take the lift of $\Theta$ to have order $n$. The distinction between
the order of a lattice automorphism and that of its action on CFT states is discussed
in~\cite{Harvey:2017rko}.

The shift must preserve the order of the orbifold generator, and its contribution to the level mismatch must
be integral after $n$ modular $T$-transformations. These requirements give
\begin{equation}
    nv\in I,
    \qquad
    \frac{n}{2}v^2\in\mathbb Z.
\end{equation}
The norm is computed with the Narain pairing, so $v^2=-|v_R|^2$. More generally, the level-matching condition
is $n(\Delta E_\Theta+v^2/2)\in\mathbb Z$, where $\Delta E_\Theta$ is the mismatch before adding the
shift~\cite{Persson:2015jka}. Under our level-matching assumption, $n\Delta E_\Theta$ is already integral,
yielding the condition above. This specialization assumes an order-$n$, level-matched lift. The additional
characteristic-vector terms appearing in the more general treatment of~\cite{Aldazabal:2025zht} must be
retained when that assumption does not hold.

To obtain an island, we must also exclude massless states in the twisted sectors. Since the right-moving
oscillators are untwisted, a massless state must have vanishing right-moving internal momentum. In the
$g$-twisted sector the momenta lie in $I^*+v$, so $v\notin I^*$ excludes such states. The same argument must
be applied to every nontrivial power of $g$. Writing $I_k=\Gamma_{22,6}^{\Theta^k}$, the momenta in the
$g^k$-twisted sector lie in $I_k^*+kv$. Since $I\subset I_k$, the projection of $I_k^*$ onto the purely
right-moving space $I\otimes\mathbb R$ is contained in $I^*$. Thus a sufficient condition for the absence of
massless states in all twisted sectors is
\begin{equation}\label{eq:shift-exact-order}
    kv\notin I^*,\qquad k=1,\ldots,n-1.
\end{equation}
Together with $nv\in I$, this states that the class of $v$ has exact order $n$ in $(I\otimes\mathbb Q)/I^*$.
We impose this condition in selecting shifts. When this sufficient condition fails in a sector, the
left-moving mass condition and the orbifold projection determine whether massless states survive.

We restrict our enumeration to shifts of exact order $n$ modulo $I^*$ and do not consider whether
shifts of smaller order can also yield islands.

\paragraph{Inequivalent shifts and the scope of the classification.}
Our analysis specifies lattice symmetries and shifts and tests their proposed S-dualities, without
explicitly constructing the full orbifold CFTs. We assume that the heterotic and type II models selected
by the conditions used here admit consistent CFT realizations. A more rigorous treatment of orbifold
CFT consistency, including the validity of the operator interpretation, is given
in~\cite{Aldazabal:2026hix}.

The allowed shifts must next be organized into equivalence classes. At the level of the invariant lattice, we identify shifts related by lattice automorphisms, translations by $I^*$, and multiplication by an integer coprime to $n$, as in \cite{Aldazabal:2025zht}. Thus the equivalence relation used in enumerating shifts is
\begin{equation}
    v_2\equiv a\sigma(v_1)\pmod{I^*},
    \qquad \sigma\in O(I),\qquad a\in\mathbb Z,\quad \gcd(a,n)=1.
\end{equation}
Multiplication by $a$ corresponds to changing the generator of the cyclic orbifold group, with the rotational part relabelled accordingly.

Two inequivalent shifts may give different presentations of the same orbifold CFT. We organize the models by
the isometry class of their perturbative electric charge lattice, whose
construction we describe below. For a given symmetry $\Theta$, this groups together all models with isometric
charge lattices. We leave open the possibility that two inequivalent shifts give rise to inequivalent CFTs
with the same charge lattice. For the order-twenty-three models, however, the one-loop partition function is
fixed by this lattice, so different massive spectra cannot distinguish shifts with isometric kernels (see Section~\ref{sec:order23} and Appendix~\ref{app:order23-partition}). For the other orders we do not
establish this stronger statement. The mismatch of shift multiplicities between the dual frames in Section~\ref{sec:order9} provides
evidence for identifications among such shift presentations.

\subsection{Perturbative electric charge lattice}
\label{ss:chargelattices}
The symmetry and shift determine the perturbative electric charges. We compute their lattice by adapting the
argument of~\cite{Persson:2015jka} to the invariant lattice of an island. The surviving six abelian gauge
fields are associated with the invariant space $I\otimes\mathbb R$. Thus the electric charge of an untwisted
state with Narain momentum $P$ is its projection
\begin{equation}
    Q=\pi_I(P),
    \qquad
    \pi_I=\frac{1}{n}\sum_{j=0}^{n-1}\Theta^j.
\end{equation}
Since $\Gamma_{22,6}$ is self-dual and $I$ is primitive, its projected lattice is
\begin{equation}
    \pi_I(\Gamma_{22,6})=I^*.
\end{equation}
The orbifold projection acts on both the momentum state and its oscillator excitations. As
in~\cite{Persson:2015jka}, the oscillator eigenvalue can compensate the phase of the momentum state without
changing its charge. Consequently, the untwisted perturbative charge lattice is $I^*$.

In the $g$-twisted sector, the charges before projection lie in $I^*+v$. The operator products of invariant
untwisted and twisted states add their charges, and the higher twisted sectors supply the corresponding
multiples of the shift. Following the same charge-generation argument as in~\cite{Persson:2015jka}, the
lattice generated by perturbative electric charges is therefore
\begin{equation}\label{eq:electric-lattice-shift}
    L_e=I^*+\mathbb Z v
    =\bigcup_{k=0}^{n-1}(I^*+kv).
\end{equation}
Unlike the presentation in~\cite{Persson:2015jka}, the shift lies in the rank-six invariant lattice itself, so
there is no separate hyperbolic plane carrying the shift.

To express this result intrinsically, define the kernel of the shift character by
\begin{equation}
    K_v\equiv\ker_I v
    :=\ker\bigl[I\longrightarrow\mathbb Q/\mathbb Z,
    \;x\longmapsto x\cdot v\bmod\mathbb Z\bigr]
    =\{x\in I\mid x\cdot v\in\mathbb Z\}.
\end{equation}
Taking the dual of~\eqref{eq:electric-lattice-shift} gives
\begin{equation}
    L_e^*
    =\{x\in I\mid x\cdot v\in\mathbb Z\}
    =K_v.
\end{equation}
Dualizing once more, we obtain
\begin{equation}\label{eq:electric-lattice-kernel}
    \boxed{L_e=(\ker_I v)^*.}
\end{equation}
The exact order of the shift enters directly in this computation. The character $x\mapsto x\cdot v\bmod\mathbb Z$ has image of order $n$, so
\begin{equation}\label{eq:exact-order-kernel-index}
 [I:K_v]=[L_e:I^*]=n,\qquad
 |\det K_v|=n^2|\det I|.
\end{equation}
In particular, the $n$ cosets in~\eqref{eq:electric-lattice-shift} are distinct. For a shift of order $d<n$
modulo $I^*$, only $d$ distinct cosets would occur, and $d$ would replace $n$ in these indices and in the
determinant formula, although $L_e=K_v^*$ would still hold. Both $K_v$ and $L_e$ have rank six and signature
$(0,6)$ in our conventions. Note that the perturbative electric charge lattice depends on the shift through
its kernel.

\subsection{Summary of the construction}

The construction can now be summarized as a classification of electric charge lattices. For each symmetry
$\Theta$ in the Leech lattice construction, we start from the corresponding rank-two invariant lattice
reported by Harada and Lang~\cite{HaradaLang}. We determine its primitive embeddings into $E_8$ and take the
orthogonal complements, with the sign of the pairing reversed, to obtain the possible rank-six invariant
lattices $I$. For each $I$, we enumerate the inequivalent shifts $v$ satisfying the modular-invariance and
twisted-sector conditions specified above. We then compute
\begin{equation}
    K_{I,v}=\ker_I v
\end{equation}
for every allowed pair $(I,v)$. We classify the theories by the isometry class of the electric charge lattice
$L_e=K_{I,v}^*$. Computing and comparing the kernels gives an equivalent way to determine this datum.

This identification can also relate constructions with different invariant lattices. A single symmetry
$\Theta$ may admit two non-isometric invariant lattices $I_1$ and $I_2$, together with allowed shifts $v_1$
and $v_2$, for which
\begin{equation}
    \ker_{I_1}v_1\simeq\ker_{I_2}v_2.
\end{equation}
The resulting charge lattices are then isometric, and we place the corresponding orbifold CFTs in the same class.

\begingroup
\small
\renewcommand{\baselinestretch}{1}\selectfont
\setlength{\tabcolsep}{10pt}
\renewcommand{\arraystretch}{1.15}
\begin{longtable}{rcccl}
    \caption[Rank-two invariant lattices in Harada--Lang]{Rank-two invariant lattices in Harada--Lang~\cite{HaradaLang}. Class names follow their convention. The last column gives the Gram matrix of $I^{\mathrm{Leech}}$. Frame shapes use the notation $\prod_{a\mid n}a^{m(a)}$ of Persson--Volpato~\cite{Persson:2015jka}, with negative exponents replacing denominators. The value of $\lambda$ is the denominator, in lowest terms, of $\frac{n}{24}\sum_{a\mid n}m(a)/a$.}\label{tab:rank-two-leech}\\
    \toprule
    No. & Class & Frame shape & $\lambda$ & $I^{\mathrm{Leech}}$ (Gram matrix) \\
    \midrule
    \endfirsthead
    \multicolumn{5}{c}{Table \thetable\ (continued)}\\
    \toprule
    No. & Class & Frame shape & $\lambda$ & $I^{\mathrm{Leech}}$ (Gram matrix) \\
    \midrule
    \endhead
    \bottomrule
    \endfoot
    1 & $-6F$ & $1^{1}\,2^{-2}\,3^{-3}\,6^{6}$ & 1 & $\left(\begin{smallmatrix}4&2\\2&4\end{smallmatrix}\right)$ \\
    2 & $8D$ & $4^{-2}\,8^{4}$ & 1 & $\left(\begin{smallmatrix}4&0\\0&4\end{smallmatrix}\right)$ \\
    3 & $9B$ & $3^{-1}\,9^{3}$ & 1 & $\left(\begin{smallmatrix}6&3\\3&6\end{smallmatrix}\right)$ \\
    4 & $-12D$ & $1^{-1}\,2^{1}\,3^{3}\,4^{-1}\,6^{-3}\,12^{3}$ & 1 & $\left(\begin{smallmatrix}4&2\\2&4\end{smallmatrix}\right)$ \\
    5 & $12G$ & $2^{-1}\,4^{2}\,6^{-1}\,12^{2}$ & 1 & $\left(\begin{smallmatrix}8&4\\4&8\end{smallmatrix}\right)$ \\
    6 & $12H$ & $1^{-1}\,2^{3}\,3^{-1}\,4^{-2}\,6^{1}\,12^{2}$ & 1 & $\left(\begin{smallmatrix}6&0\\0&6\end{smallmatrix}\right)$ \\
    7 & $-12K$ & $1^{3}\,2^{-1}\,3^{-1}\,4^{-1}\,6^{-1}\,12^{3}$ & 1 & $\left(\begin{smallmatrix}4&2\\2&4\end{smallmatrix}\right)$ \\
    8 & $12M$ & $12^{2}$ & 12 & $\left(\begin{smallmatrix}12&0\\0&12\end{smallmatrix}\right)$ \\
    9 & $-14B$ & $1^{-1}\,2^{2}\,7^{-1}\,14^{2}$ & 1 & $\left(\begin{smallmatrix}4&2\\2&8\end{smallmatrix}\right)$ \\
    10 & $15E$ & $1^{2}\,3^{-1}\,5^{-1}\,15^{2}$ & 1 & $\left(\begin{smallmatrix}4&1\\1&4\end{smallmatrix}\right)$ \\
    11 & $16A$ & $2^{2}\,4^{-1}\,8^{-1}\,16^{2}$ & 2 & $\left(\begin{smallmatrix}4&0\\0&8\end{smallmatrix}\right)$ \\
    12 & $-16B$ & $1^{2}\,2^{-1}\,8^{-1}\,16^{2}$ & 1 & $\left(\begin{smallmatrix}4&0\\0&4\end{smallmatrix}\right)$ \\
    13 & $-18B$ & $1^{2}\,2^{-1}\,3^{-1}\,9^{1}\,18^{1}$ & 1 & $\left(\begin{smallmatrix}6&3\\3&6\end{smallmatrix}\right)$ \\
    14 & $18C$ & $1^{1}\,2^{1}\,6^{-1}\,9^{-1}\,18^{2}$ & 1 & $\left(\begin{smallmatrix}4&2\\2&4\end{smallmatrix}\right)$ \\
    15 & $-18C$ & $1^{-1}\,2^{2}\,6^{-1}\,9^{1}\,18^{1}$ & 1 & $\left(\begin{smallmatrix}12&6\\6&12\end{smallmatrix}\right)$ \\
    16 & $20B$ & $4^{1}\,20^{1}$ & 4 & $\left(\begin{smallmatrix}4&0\\0&20\end{smallmatrix}\right)$ \\
    17 & $20C$ & $1^{1}\,2^{1}\,4^{-1}\,5^{-1}\,10^{1}\,20^{1}$ & 1 & $\left(\begin{smallmatrix}4&2\\2&6\end{smallmatrix}\right)$ \\
    18 & $-20C$ & $1^{-1}\,2^{2}\,4^{-1}\,5^{1}\,20^{1}$ & 1 & $\left(\begin{smallmatrix}10&0\\0&10\end{smallmatrix}\right)$ \\
    19 & $21C$ & $3^{1}\,21^{1}$ & 3 & $\left(\begin{smallmatrix}6&3\\3&12\end{smallmatrix}\right)$ \\
    20 & $22A$ & $2^{1}\,22^{1}$ & 2 & $\left(\begin{smallmatrix}6&4\\4&10\end{smallmatrix}\right)$ \\
    21 & $23A$ & $1^{1}\,23^{1}$ & 1 & $\left(\begin{smallmatrix}4&1\\1&6\end{smallmatrix}\right)$ \\
    22 & $24E$ & $2^{1}\,4^{-1}\,6^{1}\,8^{1}\,12^{-1}\,24^{1}$ & 2 & $\left(\begin{smallmatrix}4&0\\0&12\end{smallmatrix}\right)$ \\
    23 & $24F$ & $1^{1}\,3^{-1}\,4^{1}\,6^{1}\,8^{-1}\,24^{1}$ & 1 & $\left(\begin{smallmatrix}4&0\\0&6\end{smallmatrix}\right)$ \\
    24 & $-24F$ & $1^{-1}\,2^{1}\,3^{1}\,4^{1}\,8^{-1}\,24^{1}$ & 1 & $\left(\begin{smallmatrix}6&0\\0&12\end{smallmatrix}\right)$ \\
    25 & $-28A$ & $1^{1}\,2^{-1}\,4^{1}\,7^{1}\,14^{-1}\,28^{1}$ & 1 & $\left(\begin{smallmatrix}4&2\\2&8\end{smallmatrix}\right)$ \\
    26 & $-30D$ & $1^{-1}\,2^{1}\,3^{1}\,5^{1}\,15^{-1}\,30^{1}$ & 1 & $\left(\begin{smallmatrix}4&2\\2&16\end{smallmatrix}\right)$ \\
    27 & $30D$ & $1^{1}\,3^{-1}\,5^{-1}\,6^{1}\,10^{1}\,15^{1}$ & 1 & $\left(\begin{smallmatrix}8&2\\2&8\end{smallmatrix}\right)$ \\
    28 & $-30E$ & $2^{1}\,3^{1}\,5^{1}\,6^{-1}\,10^{-1}\,30^{1}$ & 1 & $\left(\begin{smallmatrix}4&1\\1&4\end{smallmatrix}\right)$ \\
\end{longtable}
\endgroup

\section{Construction of type II islands}
\label{sec:typeII-construction}

The type II construction follows the same lattice procedure, with an additional condition on the action on
spacetime spinors. We start with type II strings on $T^6$, whose Narain lattice is $\Gamma_{6,6}$. We take
$\Theta$ to act on all six left-moving internal directions and trivially on the right movers. The invariant
and coinvariant lattices therefore have signatures $(0,6)$ and $(6,0)$, respectively. The lattice $N$ may
contain roots, since these do not give massless vector multiplets in the type II string. A further condition
is required on the action on spacetime spinors. Starting from thirty-two supercharges, we must retain
precisely the sixteen right-moving supercharges and project out all sixteen left-moving ones.

\subsection{Lattice symmetries and their integral realizations}

The required lattice symmetries are obtained from $E_8$, which plays the role of the Leech lattice in the
heterotic construction. Choose $\Theta'\in O(E_8)=W(E_8)$ with a rank-two fixed lattice $F$, and let
\begin{equation}
    F=E_8^{\Theta'},\qquad N=F^\perp\subset E_8.
\end{equation}
We use Carter's classification and notation for conjugacy classes in the Weyl group~\cite{Carter}. Each class
admits a representative expressed as a product of reflections in linearly independent roots, and the class
label refers to this reflection presentation. For example, $6A_1$ denotes a product of reflections in six
mutually orthogonal roots. Labels such as $D_4(a_1)$ distinguish further conjugacy classes in Carter's
notation. To obtain the lattice used in our construction, we take the saturation of the span of the reflecting
roots in $E_8$.

Let $\alpha_1,\ldots,\alpha_6$ be the reflecting roots in such a presentation, and write
\begin{equation}
    R_{\mathrm C}=\sum_{i=1}^{6}\mathbb Z\alpha_i.
\end{equation}
Their rational span is the rotated subspace. The lattice entering our construction is its full intersection with $E_8$, namely the saturation
\begin{equation}\label{eq:Carter-saturation}
    N=\operatorname{Sat}_{E_8}(R_{\mathrm C})
      =(R_{\mathrm C}\otimes\mathbb Q)\cap E_8
      =(E_8^{\Theta'})^\perp_{E_8}.
\end{equation}
In particular, $R_{\mathrm C}\subseteq N$ can have nontrivial finite index. Table~\ref{tab:Carter-saturations}
specifies the saturated lattice used for every Carter class appearing in our construction. Here $A_r,D_r,E_r$
denote root lattices with roots of squared length two. For example, $6A_1$ gives $N=D_6$ with index four, and
$3A_2$ gives $N=E_6$ with index three. Different Carter classes can therefore act on the same lattice $N$,
while defining different symmetries.

\begin{table}[htbp]
    \centering
    \small
    \renewcommand{\arraystretch}{1.15}
    \begin{tabular}{lccc}
        \toprule
        Carter class & Reflecting-root lattice $R_{\mathrm C}$ & Saturated $N$ & $[N:R_{\mathrm C}]$ \\
        \midrule
        $6A_1$ & $6A_1$ & $D_6$ & 4 \\
        $3A_2$ & $3A_2$ & $E_6$ & 3 \\
        $A_3+3A_1$ & $A_3+3A_1$ & $D_5+A_1$ & 2 \\
        $A_2+4A_1$ & $A_2+4A_1$ & $D_4+A_2$ & 2 \\
        $D_4+2A_1$ & $D_4+2A_1$ & $D_6$ & 2 \\
        $2A_2+2A_1$ & $2A_2+2A_1$ & $2A_2+2A_1$ & 1 \\
        $D_4(a_1)+A_2$ & $D_4+A_2$ & $D_4+A_2$ & 1 \\
        $D_6,\ D_6(a_1),\ D_6(a_2)$ & $D_6$ & $D_6$ & 1 \\
        $E_6,\ E_6(a_1),\ E_6(a_2)$ & $E_6$ & $E_6$ & 1 \\
        $A_6$ & $A_6$ & $A_6$ & 1 \\
        $A_4+2A_1$ & $A_4+2A_1$ & $A_4+2A_1$ & 1 \\
        $A_3+A_2+A_1$ & $A_3+A_2+A_1$ & $A_3+A_2+A_1$ & 1 \\
        $D_5(a_1)+A_1$ & $D_5+A_1$ & $D_5+A_1$ & 1 \\
        $A_4+A_2$ & $A_4+A_2$ & $A_4+A_2$ & 1 \\
        \bottomrule
    \end{tabular}
    \caption{Carter classes and the saturated coinvariant lattices used in the type II construction. A plus sign between lattice types denotes an orthogonal direct sum. The saturation is taken in the corresponding embedding in $E_8$. Different Spin lifts of the same lattice symmetry share the same entry.}
    \label{tab:Carter-saturations}
\end{table}

The lattice $N$ has rank six, and the restriction of $\Theta'$ acts trivially on $A_N$. As before, for a
primitive embedding of $F$ into $E_8$ with complement $I^{+}$,
\begin{equation}
    q_N\simeq-q_F\simeq q_{I^{+}}.
\end{equation}
Gluing $N$ to $I=I^{+}(-1)$ therefore produces $\Gamma_{6,6}$. We determine the possible complements
by the same procedure as in the heterotic case, with the same restriction that possible gluing
degeneracies at fixed $(N,I)$ are not considered.

We restrict here to the saturated coinvariant lattices $N$ obtained from $E_8$. Further integral
realizations of the same real action may arise from proper finite-index sublattices $N'\subset N$,
provided that $\Theta$ preserves $N'$, acts trivially on its discriminant group, and $N'$ admits a primitive
embedding into $\Gamma_{6,6}$. We do not analyse these possibilities, so our construction is not claimed
to exhaust the type II islands.

\subsection{Spin lifts and shifts}

The action on Ramond states requires a choice of Spin lift of the lattice automorphism. For each
orientation-preserving left-moving rotation we must examine its two lifts to $\operatorname{Spin}(6)$. Write
the vector eigenvalues as $e^{\pm2\pi i\phi_a}$, $a=1,2,3$. The eigenvalues on a chiral spinor are
\begin{equation}
    \eta\exp\!\left(\pi i\sum_{a=1}^{3}\epsilon_a\phi_a\right),
    \qquad \eta=\pm1,\quad \epsilon_a=\pm1,
\end{equation}
with the parity of the number of minus signs fixed by the spinor chirality. We retain the lifts with no
invariant left-moving supercharge and take the right-moving lift to be the identity. The order of the lift can
differ from that of the vector action, in particular by a factor of two. In this section $n$ denotes the order
of the lifted symmetry, which is the order relevant to the orbifold construction.

For each such lift, we enumerate shifts along $I$, imposing the order and modular-invariance conditions and
excluding additional massless states in all twisted sectors. The procedure parallels the heterotic analysis,
but the level mismatch must be computed with both worldsheet bosons and fermions and with the chosen lift. The
massless-spectrum check must also exclude any restoration of the projected supercharges in twisted sectors.

When the lifted symmetry is level matched before adding the shift and the same order-$n$ shift assumptions
apply, the perturbative NSNS electric charge lattice is obtained by the same projection and kernel argument:
\begin{equation}
    L_{e,\mathrm{pert}}=I^*+\mathbb Z v=(\ker_I v)^*.
\end{equation}
We again classify these constructions by their electric charge lattices, using the dual kernels to compare isometry classes.

\section{S-duality}
\label{sec:s-duality}

In this section we formulate tests of S-duality for the heterotic and type II islands constructed above.
We first review the duality chains available for CHL models, then propose an extension of their gravitational
thresholds to islands. The resulting frame shapes identify candidate dual descriptions under Fricke and
Atkin--Lehner transformations. We then formulate the corresponding tests on electromagnetic charge lattices,
which distinguish individual islands within each archipelago. The explicit results are presented in
Section~\ref{sec:explicit-results}.

\subsection{S-duality in CHL models}
\label{sec:chl-duality}

Within the cyclic constructions considered here, the CHL models of \cite{Persson:2015jka} are precisely the
four-dimensional models with nonzero vector-multiplet gauge-group rank, $n_v>0$. The string islands instead
have $n_v=0$. This distinction separates the CHL duality analysis that we review below from its proposed
extension to islands.

The moduli spaces and heterotic--type IIA dual descriptions of CHL models were studied
in~\cite{Chaudhuri:1995bf,Chaudhuri:1995dj,Chaudhuri:1995ee}. We review the duality argument
of~\cite{Gregori:1997hi,Persson:2015jka}, emphasizing the role of the spectator torus and the type IIA frame.

We first consider heterotic constructions for which the coinvariant lattice $N$ embeds primitively into the
Leech lattice and has rank at most twenty. Such coinvariant lattices admit primitive embeddings into
$\Gamma_{20,4}$~\cite{Persson:2015jka}. The relation between supersymmetry-preserving K3 sigma-model
symmetries and the Conway group was established in~\cite{Gaberdiel:2011fg}. Together with the uniqueness of
the embedding after taking the orthogonal direct sum with spectator hyperbolic planes, this allows a
compatible polarization
\begin{equation}
    \Gamma_{22,6}\simeq\Gamma_{20,4}\oplus\Gamma_{2,2},
    \qquad
    \Theta=\Theta_{20,4}\oplus 1.
\end{equation}
The symmetry therefore admits a presentation in heterotic string theory on $T^4\times T^2$, acting only on the
internal theory associated with $T^4$. We can choose the accompanying shift to be a translation along a circle
of the spectator $T^2$. This can be represented by $v=u/n$, where $u$ is a primitive null vector in
$\Gamma_{2,2}$ and $n$ is the order of the symmetry. The CHL presentation uses this particular class of
shifts.

The six-dimensional duality between the heterotic string on $T^4$ and type IIA on K3 can then be applied by
the adiabatic argument~\cite{Persson:2015jka}. The resulting description is type IIA on
\begin{equation}
    (K3\times T^2)/\langle g\rangle,
    \qquad g=(\Theta_{K3},\delta),
\end{equation}
where $\Theta_{K3}$ is the image of the heterotic symmetry in the K3 sigma model and $\delta$ is the
translation along the spectator circle. The symbol $\Theta_{K3}$ denotes a symmetry of the sigma model and
need not be a geometric automorphism of K3. It preserves the supersymmetry of the six-dimensional theory, so
the quotient has sixteen supercharges in four dimensions.

This frame makes the four-dimensional S-duality problem accessible to perturbation theory. Heterotic--type IIA
duality exchanges the heterotic axiodilaton with the K\"ahler modulus of the type IIA $T^2$. Consequently,
S-duality in the heterotic description is realized as T-duality on the spectator torus in the type IIA
description. One can determine the transformed orbifold by perturbative T-duality and then use the duality
chain to identify its interpretation in the original variables. Persson and Volpato found that, for the models
they studied, the resulting descriptions are again heterotic or type II asymmetric orbifolds with all sixteen
supercharges arising from one worldsheet sector~\cite{Persson:2015jka}.

An analogous argument applies to the type II constructions with coinvariant lattice of rank at most four,
using a presentation in which the rotation is supported on $\Gamma_{4,4}$ and a circle of the spectator $T^2$
carries the shift. Six-dimensional string--string duality maps the asymmetric action, which preserves all
right-moving supercharges and projects out all left-moving ones, to a type IIA description on $T^4\times T^2$
in which the rotation projects out half of the left-moving and half of the right-moving supercharges. Thus the
sixteen surviving four-dimensional supercharges are divided equally between the two sectors. This is a $(2,2)$
description, where each entry counts four real spacetime supercharges, in contrast to the original $(0,4)$
description. Perturbative torus duality and the corresponding duality chain then allow us to identify the
candidate S-duals.

These are the duality chains studied by Persson and Volpato~\cite{Persson:2015jka}. Their essential ingredient
is the spectator $T^2$, together with a dual $(2,2)$ description in which the S-duality of interest becomes a
perturbative T-duality. The rank bounds above explain why this argument is available for these constructions.
The full family of heterotic and type II models constructed from these coinvariant lattices, with the
spectator-circle shifts described above, closes under the S-duality operations considered by Persson and
Volpato, with every proposed dual again belonging to this family.

\subsection{\texorpdfstring{$R^2$}{R2} thresholds and frame shapes}

The gravitational threshold is the moduli-dependent coefficient of a four-derivative term in the low-energy
effective action. Following~\cite{Gregori:1997hi,Persson:2015jka}, we use $R^2$ as shorthand for the CP-even
curvature-squared coupling, represented schematically by
\begin{equation}
 \int d^4x\,\sqrt{-g}\,f_{R^2}(S,\bar S)
 R_{\mu\nu\rho\sigma}R^{\mu\nu\rho\sigma},
\end{equation}
with its supersymmetric completion understood. Ricci-tensor and scalar-curvature terms can be rearranged by
field redefinitions. A common representative uses the Gauss--Bonnet combination
$R_{\mu\nu\rho\sigma}^2-4R_{\mu\nu}^2+R^2$. The threshold discussed here is the moduli dependence of this
gravitational coupling. It is BPS-saturated because the relevant helicity supertrace cancels contributions from long
supersymmetry multiplets, leaving the protected contribution of short multiplets. This property makes it a
useful observable for comparing dual descriptions.

The protected $R^2$ coupling provides a way to follow the S-dualities of the CHL models. In the $(2,2)$
description, the threshold is computed at one loop in terms of the K\"ahler modulus of the spectator torus. In
the K3 frame, the symmetry data enter through an eta product specified by a frame
shape~\cite{Gregori:1997hi,Persson:2015jka}. Translating this result to the $(0,4)$ frame replaces the torus
modulus by the axiodilaton $S$, giving a coupling whose moduli dependence is entirely through $S$. It includes
contributions that are nonperturbative in this frame.

Write the frame shape, its eta product and the modular weight as
\begin{equation}
    \pi=\prod_{a\mid n}a^{m(a)},\qquad
    \eta_\pi(S)=\prod_{a\mid n}\eta(aS)^{m(a)},\qquad
    k=\frac12\sum_{a\mid n}m(a).
\end{equation}
In the normalization in which Fricke duality acts as $S\mapsto-1/(nS)$, the CP-even threshold takes the form
\begin{equation}\label{eq:CHL-R2-threshold}
    f_{R^2}^{\pi}(S,\bar S)
    =-\log\!\left[S_2^{2k}|\eta_\pi(S)|^4\right]
      +\text{const},\qquad S_2=\operatorname{Im}S.
\end{equation}
We suppress an overall convention-dependent normalization of the $R^2$ operator. The nonholomorphic factor
$S_2^{2k}$ compensates the weight of the eta product, while the additive constant depends on the subtraction
convention.

For a heterotic CHL model, the frame shape in this formula is determined by the left-moving action of the
Narain lattice symmetry. More precisely, let $V_L^{24}$ be the space of the 22 internal left-moving bosonic
oscillator directions together with the two unrotated transverse directions in light-cone gauge. The frame
shape encodes the eigenvalues on this space through
\begin{equation}
    \det\!\left(x-\Theta\big|_{V_L^{24}}\right)=\prod_{a\mid n}(x^a-1)^{m(a)}.
\end{equation}
The notation $\prod a^{m(a)}$ is a formal label for this polynomial factorization. Each factor $x^a-1$
supplies the $a$-th roots of unity, with cancellations when some exponents are negative. The six right-moving
Narain directions are excluded from this 24-dimensional action.
Thus the eigenvalues of $\Theta$ directly determine the eta product. For type II CHL models in the $(0,4)$
frame, the threshold frame shape can be recovered from the action on spinors. We now describe this
prescription and propose its extension to type II islands.

\subsection{Type II threshold frame shapes}

We refer to the frame shape specifying the eta quotient in the gravitational threshold as its \emph{threshold
frame shape}. For type II models in a $(0,4)$ description, the following prescription extracts its exponents
from the Spin action. It reproduces the known CHL thresholds and supplies the conjectured thresholds for
islands discussed below.

In this subsection, $g$ denotes the action of the chosen type II Spin lift on the supercharges. The
accompanying translation acts trivially on them. A Spin lift is a preimage of the internal rotation under the
double cover $\operatorname{Spin}(6)\to SO(6)$. We preserve all sixteen right-moving supercharges and project
out all sixteen left-moving ones. Write the latter as $Q_\alpha^A$ and their Hermitian conjugates, where
$\alpha=1,2$ is a four-dimensional Weyl-spinor index and $A=1,\ldots,4$ is the internal index. These give
sixteen real supercharges. The internal multiplicity space is the complex chiral spinor representation
$S_{\mathrm{br}}\cong\mathbf4$ of $\operatorname{Spin}(6)\cong SU(4)$. The determinant below acts only on this
four-dimensional internal space. The quantity we need is
\begin{equation}\label{eq:spinor-determinant}
    \nu_g(r)=\det\!\left(1-g^r\big|_{S_{\mathrm{br}}}\right)
       =\prod_{\alpha=1}^{4}(1-\lambda_\alpha^r),
\end{equation}
where $\lambda_\alpha$ are the eigenvalues of the chosen lift on $S_{\mathrm{br}}$. This gives an intrinsic
prescription in the $(0,4)$ frame, using only the action on the supercharges.

Let $n$ be the order of the lift. We recover the frame-shape exponents from
\begin{equation}\label{eq:spinor-divisor-character}
    \nu_g(r)=\sum_{a\mid\gcd(r,n)}a\,m(a),
\end{equation}
by M\"obius inversion:
\begin{equation}\label{eq:spinor-frame-exponents}
    m(a)=\frac1a\sum_{d\mid a}\mu(a/d)\,\nu_g(d),
    \qquad a\mid n.
\end{equation}
Here $\mu$ is the M\"obius function. For the symmetries considered here, these exponents are integral and
satisfy~\eqref{eq:spinor-divisor-character} for every power of $g$. They define the \emph{threshold frame
shape} and its eta quotient,
\begin{equation}
    \pi_g=\prod_{a\mid n}a^{m(a)},\qquad
    \eta_g(S)=\prod_{a\mid n}\eta(aS)^{m(a)}.
\end{equation}
The exponents may be negative. Since $g^n=1$, the determinant vanishes at $r=n$, so $\sum_{a\mid n}a\,m(a)=0$;
the unweighted sum instead determines the modular weight.

For every type II asymmetric orbifold in the CHL family, this determinant prescription reproduces precisely
the frame shape entering its known $R^2$ threshold. The comparison with the thresholds obtained through the
$(2,2)$ descriptions therefore identifies their eta products directly from the spinorial action in the $(0,4)$
frame. This agreement is the basis for extending the prescription to islands, where that dual description is
unavailable.

\paragraph{Example: $\boldsymbol{g=(-1)^{F_L}}$.}
Here $F_L$ is left-moving spacetime fermion number. The symmetry acts as $-\mathbf1_4$ on $S_{\mathrm{br}}$, giving
\begin{equation}
    \nu_g(r)=\bigl(1-(-1)^r\bigr)^4
    =\begin{cases}
        16,&r\ {\rm odd},\\
        0,&r\ {\rm even}.
      \end{cases}
\end{equation}
Thus $m(1)=16$ and $m(1)+2m(2)=0$, and hence
\begin{equation}
    \pi_g=1^{16}2^{-8},\qquad
    \eta_g(S)=\frac{\eta(S)^{16}}{\eta(2S)^8}.
\end{equation}
This is the eta quotient appearing in the CHL threshold for this type II symmetry.

\paragraph{Extension to islands.}
We conjecture that the same determinant prescription gives the eta quotient of the $R^2$ threshold for type II
islands. The conjecture concerns the identification with the gravitational coupling. The exponents themselves
are computed directly from the Spin lift using~\eqref{eq:spinor-frame-exponents}. For heterotic islands, we
likewise assume that the threshold retains~\eqref{eq:CHL-R2-threshold}, with the ordinary frame shape of the
Narain lattice symmetry. Together, these assumptions specify the $R^2$ thresholds of all the heterotic and
type II islands considered here. These are predictions for the full nonperturbative couplings. An explicit
calculation is not currently available for islands, and our test is the agreement between the duality
assignments implied by this ansatz and the independently computed electromagnetic charge lattices.

Table~\ref{tab:typeII-threshold-frames} records the type II symmetries giving islands and their threshold
frame shapes. Its Carter labels specify the reflection classes, and the corresponding saturated lattices $N$ are
listed separately in Table~\ref{tab:Carter-saturations}. The order $n$ includes the Spin lift, as in the
preceding discussion. The two lifts of $D_4(a_1)+A_2$ both have order twelve and are distinguished by the
labels I and II.

\begingroup
\small
\renewcommand{\baselinestretch}{1}\selectfont
\setlength{\tabcolsep}{10pt}
\renewcommand{\arraystretch}{1.45}
\begin{longtable}{lcl}
    \caption{Type II island symmetries in Carter notation, their Spin-lift orders, and their intrinsic $(0,4)$ threshold frame shapes.}\label{tab:typeII-threshold-frames}\\
    \toprule
    Carter class & Spin-lift order $n$ & Threshold frame shape \\
    \midrule
    \endfirsthead
    \multicolumn{3}{c}{Table \thetable\ (continued)}\\
    \toprule
    Carter class & Spin-lift order $n$ & Threshold frame shape \\
    \midrule
    \endhead
    \bottomrule
    \endfoot
    $6A_1$ & 4 & $1^{-4}\,2^{10}\,4^{-4}$ \\
    $3A_2$ & 6 & $1^{-2}\,2\,3^{6}\,6^{-3}$ \\
    $A_2+4A_1$ & 6 & $1^{-3}\,2^{6}\,3\,6^{-2}$ \\
    $E_6(a_2)$ & 6 & $1^{6}\,2^{-3}\,3^{-2}\,6$ \\
    $A_3+3A_1$ & 8 & $1^{-2}\,2^{3}\,4^{3}\,8^{-2}$ \\
    $D_6(a_1)$ & 8 & $1^{4}\,2^{-2}$ \\
    $E_6(a_1)$ & 9 & $1^{3}\,3^{-1}$ \\
    $D_4+2A_1$ & 12 & $1^{-1}\,2\,3^{-1}\,4^{2}\,6^{3}\,12^{-2}$ \\
    $2A_2+2A_1$ & 12 & $1^{-2}\,2^{3}\,3^{2}\,4^{-1}\,6\,12^{-1}$ \\
    $D_6(a_2)$ & 12 & $1^{2}\,2\,3^{-2}\,4^{-1}\,6^{3}\,12^{-1}$ \\
    $D_4(a_1)+A_2\text{ (lift I)}$ & 12 & $1^{-1}\,2^{-1}\,3^{3}\,4^{3}\,6^{-1}\,12^{-1}$ \\
    $D_4(a_1)+A_2\text{ (lift II)}$ & 12 & $1^{3}\,2^{-3}\,3^{-1}\,4^{3}\,6\,12^{-1}$ \\
    $E_6$ & 12 & $1^{2}\,2^{-1}\,3^{2}\,6^{-1}$ \\
    $A_6$ & 14 & $1^{2}\,2^{-1}\,7^{2}\,14^{-1}$ \\
    $A_4+A_2$ & 15 & $1^{-1}\,3^{2}\,5^{2}\,15^{-1}$ \\
    $E_6(a_1)$ & 18 & $1\,2\,3^{-1}\,9^{2}\,18^{-1}$ \\
    $A_4+2A_1$ & 20 & $1^{-1}\,2\,4\,5\,10\,20^{-1}$ \\
    $D_6$ & 20 & $1\,4\,5^{-1}\,10^{2}\,20^{-1}$ \\
    $A_3+A_2+A_1$ & 24 & $1^{-1}\,2\,3\,8\,12\,24^{-1}$ \\
    $D_5(a_1)+A_1$ & 24 & $1\,3^{-1}\,6\,8\,12\,24^{-1}$ \\
    $A_4+A_2$ & 30 & $1\,2^{-1}\,6\,10\,15\,30^{-1}$ \\
\end{longtable}
\endgroup

\subsection{Closure of the frame shapes under S-duality}

With these conjectured thresholds, S-duality acts on the frame shapes by the same rules as in the CHL models.
In the formulas below, $\eta_\Theta$ denotes the eta product assigned to the symmetry, using its ordinary
frame shape for a heterotic model and its threshold frame shape for a type II model.

Applying the Fricke operation and using $\eta(-1/\tau)=(-i\tau)^{1/2}\eta(\tau)$ gives
\begin{equation}
    \eta_\Theta\!\left(-\frac{1}{nS}\right)
    =(-iS)^k\prod_{a\mid n}\left(\frac{n}{a}\right)^{m(a)/2}
    \prod_{a\mid n}\eta\!\left(\frac{n}{a}S\right)^{m(a)}.
\end{equation}
The factor $|S|^{4k}$ cancels against the transformation of $S_2^{2k}$. The remaining $S$-dependent expression
is again of the form~\eqref{eq:CHL-R2-threshold}, but with
\begin{equation}\label{eq:Fricke-frame-shape}
    m'(a)=m(n/a),\qquad
    \pi_{\Theta'}=\prod_{a\mid n}a^{m(n/a)}.
\end{equation}
Consequently,
\begin{equation}
    f_{R^2}^{\Theta}\!\left(-\frac{1}{nS},-\frac{1}{n\bar S}\right)
    =f_{R^2}^{\Theta'}(S,\bar S)+\text{const}.
\end{equation}
If the frame shape is balanced, $m(a)=m(n/a)$, the threshold is invariant up to a constant. Otherwise it transforms into a different eta-product threshold.

More generally, the Atkin--Lehner operations considered in~\cite{Persson:2015jka,Persson:2017lkn} similarly
permute the exponents. For an exact divisor $e$ of $n$, their rule is
\begin{equation}
    m_e(a)=m\!\left(\frac{ea}{\gcd(a,e)^2}\right),
    \qquad e\mid n,\quad \gcd(e,n/e)=1.
\end{equation}
The preservation of the functional form suggests an asymmetric-orbifold interpretation of the dual theory and
specifies the eigenvalue data to compare. The integral lattice and the shift are additional data needed to
specify the orbifold. In the CHL analysis, its transformation agrees with the dual models obtained through the
duality chains described above. It therefore provides a concrete observable through which those
identifications can be tested.

The key result is that the complete set of frame shapes obtained for our heterotic and type II islands closes
under these operations. We find that every transformed frame shape
occurs again among the constructions. Conditional on the threshold conjecture, their $R^2$ corrections
therefore form a closed set under the S-duality operations of Persson and Volpato. This identifies candidate
dual archipelagos. Determining how their individual islands are paired requires the charge-lattice analysis
below.

\begin{figure}[htbp]
    \centering
    \begin{minipage}[t]{.48\textwidth}
        \centering
        $n=6$\par\smallskip
        \resizebox{\linewidth}{!}{%
        \begin{tikzpicture}[font=\small,>=stealth,
            every node/.style={inner sep=3pt,text=black},every path/.style={thin,draw=blue}]
        \node (A) at (0,0) {$\mathrm{HET}(-6F)$};
        \node (B) at (4.5,0) {$\mathrm{II}(E_6(a_2))$};
        \node (C) at (0,-1.6) {$\mathrm{II}(3A_2)$};
        \node (D) at (4.5,-1.6) {$\mathrm{II}(A_2+4A_1)$};
        \draw[<->] (A) -- node[above,text=blue] {$W_6$} (B);
        \draw[<->] (C) -- node[below,text=blue] {$W_6$} (D);
        \draw[<->] (A) -- node[left,text=blue] {$W_2$} (C);
        \draw[<->] (B) -- node[right,text=blue] {$W_2$} (D);
        \draw[<->] (A.south east) -- (D.north west);
        \draw[<->] (C.north east) -- (B.south west);
        \node[text=blue,fill=white,inner sep=1pt] at (2.15,-.46) {$W_3$};
        \node[text=blue,fill=white,inner sep=1pt] at (3.05,-.88) {$W_3$};
        \end{tikzpicture}%
        }
    \end{minipage}\hfill
    \begin{minipage}[t]{.48\textwidth}
        \centering
        $n=12$\par\smallskip
        \resizebox{\linewidth}{!}{%
        \begin{tikzpicture}[font=\small,>=stealth,
            every node/.style={inner sep=3pt,text=black},every path/.style={thin,draw=blue}]
        \node (A) at (0,0) {$\mathrm{HET}(12H)$};
        \node (B) at (4.5,0) {$\mathrm{II}(D_6(a_2))$};
        \node (C) at (0,-1.6) {$\mathrm{II}(2A_2+2A_1)$};
        \node (D) at (4.5,-1.6) {$\mathrm{II}(D_4+2A_1)$};
        \draw[<->] (A) -- node[above,text=blue] {$W_{12}$} (B);
        \draw[<->] (C) -- node[below,text=blue] {$W_{12}$} (D);
        \draw[<->] (A) -- node[left,text=blue] {$W_4$} (C);
        \draw[<->] (B) -- node[right,text=blue] {$W_4$} (D);
        \draw[<->] (A.south east) -- (D.north west);
        \draw[<->] (C.north east) -- (B.south west);
        \node[text=blue,fill=white,inner sep=1pt] at (2.15,-.46) {$W_3$};
        \node[text=blue,fill=white,inner sep=1pt] at (3.05,-.88) {$W_3$};
        \end{tikzpicture}%
        }
    \end{minipage}
    \par\medskip
    \begin{minipage}[t]{.48\textwidth}
        \centering
        $n=18$\par\smallskip
        \resizebox{\linewidth}{!}{%
        \begin{tikzpicture}[font=\small,>=stealth,
            every node/.style={inner sep=3pt,text=black},every path/.style={thin,draw=blue}]
        \node (A) at (0,0) {$\mathrm{HET}(-18B)$};
        \node (B) at (4.5,0) {$\mathrm{HET}(18C)$};
        \node (C) at (0,-1.6) {$\mathrm{HET}(-18C)$};
        \node (D) at (4.5,-1.6) {$\mathrm{II}(E_6(a_1))$};
        \draw[<->] (A) -- node[above,text=blue] {$W_{18}$} (B);
        \draw[<->] (C) -- node[below,text=blue] {$W_{18}$} (D);
        \draw[<->] (A) -- node[left,text=blue] {$W_2$} (C);
        \draw[<->] (B) -- node[right,text=blue] {$W_2$} (D);
        \draw[<->] (A.south east) -- (D.north west);
        \draw[<->] (C.north east) -- (B.south west);
        \node[text=blue,fill=white,inner sep=1pt] at (2.15,-.46) {$W_9$};
        \node[text=blue,fill=white,inner sep=1pt] at (3.05,-.88) {$W_9$};
        \end{tikzpicture}%
        }
    \end{minipage}\hfill
    \begin{minipage}[t]{.48\textwidth}
        \centering
        $n=30$\par\smallskip
        \resizebox{\linewidth}{!}{%
        \begin{tikzpicture}[font=\small,>=stealth,
            every node/.style={inner sep=3pt,text=black},every path/.style={thin,draw=blue}]
        \node (A) at (0,0) {$\mathrm{HET}(30D)$};
        \node (B) at (4.5,0) {$\mathrm{HET}(-30D)$};
        \node (C) at (0,-1.6) {$\mathrm{HET}(-30E)$};
        \node (D) at (4.5,-1.6) {$\mathrm{II}(A_4+A_2)$};
        \draw[<->] (A) -- node[above,text=blue] {$W_5,W_3$} (B);
        \draw[<->] (C) -- node[below,text=blue] {$W_5,W_3$} (D);
        \draw[<->] (A) -- node[left,text=blue] {$W_{30},W_2$} (C);
        \draw[<->] (B) -- node[right,text=blue] {$W_{30},W_2$} (D);
        \draw[<->] (A.south east) -- (D.north west);
        \draw[<->] (C.north east) -- (B.south west);
        \node[text=blue,fill=white,inner sep=1pt] at (2.15,-.46) {$W_{10},W_6$};
        \node[text=blue,fill=white,inner sep=1pt] at (3.05,-.88) {$W_{10},W_6$};
        \draw[->] (A.north) .. controls +(-.45,.65) and +(.45,.65) .. node[above,text=blue] {$W_{15}$} (A.north);
        \draw[->] (B.north) .. controls +(-.45,.65) and +(.45,.65) .. node[above,text=blue] {$W_{15}$} (B.north);
        \draw[->] (C.south) .. controls +(-.45,-.65) and +(.45,-.65) .. node[below,text=blue] {$W_{15}$} (C.south);
        \draw[->] (D.south) .. controls +(-.45,-.65) and +(.45,-.65) .. node[below,text=blue] {$W_{15}$} (D.south);
        \end{tikzpicture}%
        }
    \end{minipage}
    \caption{The four-node frame-shape orbits. Each arrow represents the indicated Atkin--Lehner operation. The order-thirty orbit also has $W_{15}$ loops at every node.}
    \label{fig:orbits-four-node}
\end{figure}
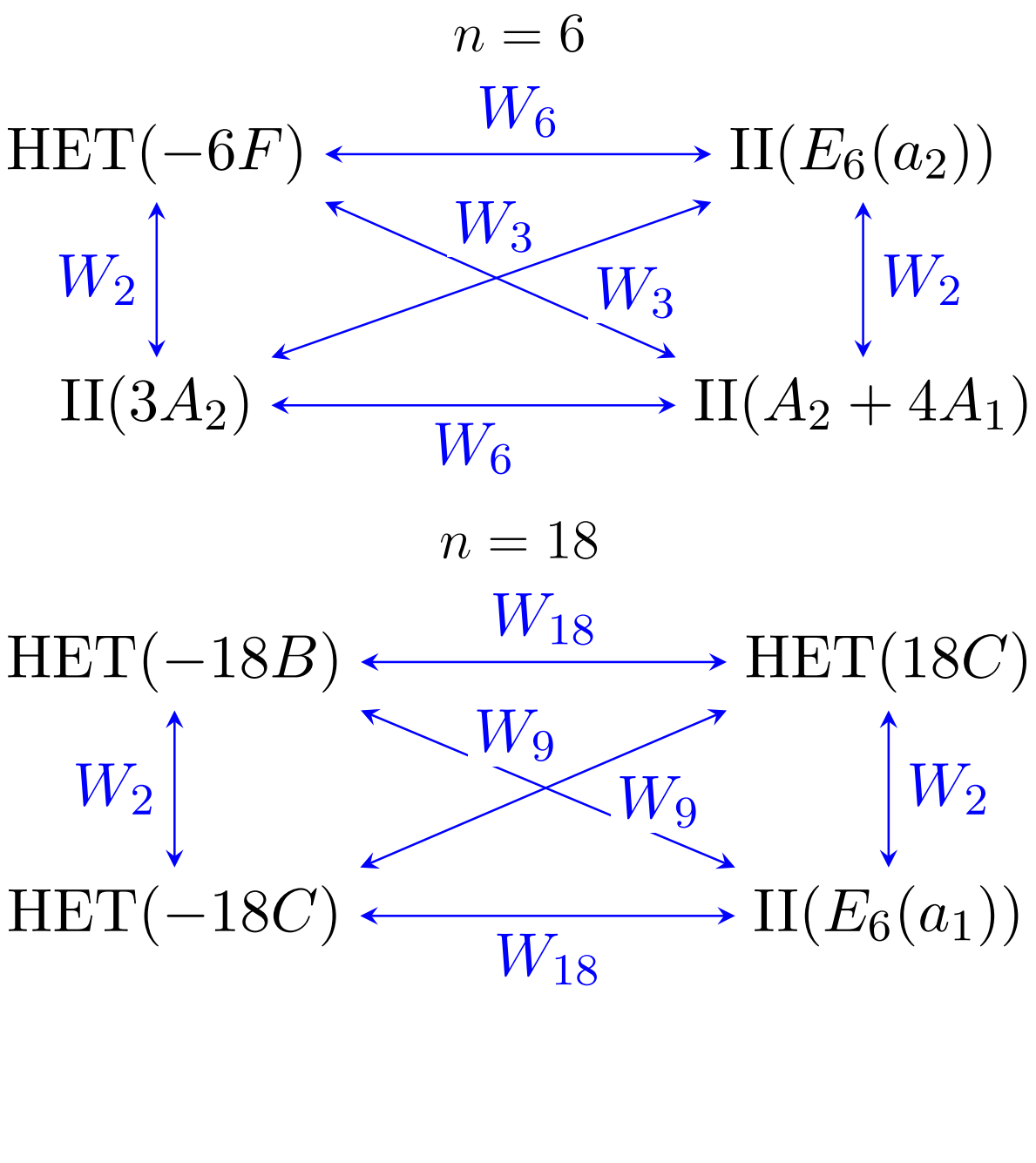

The frame-shape orbits are summarized in Figure~\ref{fig:orbits-four-node} and Table~\ref{tab:orbits-small}.
We retain diagrams for the four-node orbits and list the two-node and fixed orbits in a table. A node
$\mathrm{HET}(X)$ denotes the archipelago associated with Conway class $X$, and $\mathrm{II}(Y)$ denotes the
one associated with Carter class $Y$ and the specified Spin-lift order $n$. Edges are labelled by the
Atkin--Lehner operation $W_e$. At a fixed node the frame shape is invariant, while S-duality may permute the
individual islands. The two occurrences of $D_4(a_1)+A_2$ represent distinct Spin lifts, and the shared label
$23A/23B$ denotes the same frame shape.

\begin{table}[htbp]
    \centering
    \small
    \renewcommand{\baselinestretch}{1}\selectfont
    \setlength{\tabcolsep}{5pt}
    \renewcommand{\arraystretch}{1.25}
    \begin{tabular}{clcc}
        \toprule
        $n$ & Symmetry descriptions & \shortstack{Exchange\\descriptions} & \shortstack{Preserve each\\description} \\
        \midrule
        8 & $\mathrm{HET}(8D),\ \mathrm{II}(D_6(a_1))$ & $W_8$ & --- \\
        9 & $\mathrm{HET}(9B),\ \mathrm{II}(E_6(a_1))$ & $W_9$ & --- \\
        12 & $\mathrm{HET}(-12D),\ \mathrm{II}(D_4(a_1)+A_2)$ & $W_{12},W_3$ & $W_4$ \\
        12 & $\mathrm{HET}(12G),\ \mathrm{II}(E_6)$ & $W_{12},W_4$ & $W_3$ \\
        12 & $\mathrm{HET}(-12K),\ \mathrm{II}(D_4(a_1)+A_2)$ & $W_4,W_3$ & $W_{12}$ \\
        14 & $\mathrm{HET}(-14B),\ \mathrm{II}(A_6)$ & $W_{14},W_2$ & $W_7$ \\
        15 & $\mathrm{HET}(15E),\ \mathrm{II}(A_4+A_2)$ & $W_5,W_3$ & $W_{15}$ \\
        20 & $\mathrm{HET}(20C),\ \mathrm{II}(A_4+2A_1)$ & $W_5,W_4$ & $W_{20}$ \\
        20 & $\mathrm{HET}(-20C),\ \mathrm{II}(D_6)$ & $W_{20},W_5$ & $W_4$ \\
        24 & $\mathrm{HET}(24F),\ \mathrm{II}(A_3+A_2+A_1)$ & $W_8,W_3$ & $W_{24}$ \\
        24 & $\mathrm{HET}(-24F),\ \mathrm{II}(D_5(a_1)+A_1)$ & $W_{24},W_3$ & $W_8$ \\
        \midrule
        4 & $\mathrm{II}(6A_1)$ & --- & $W_4$ \\
        8 & $\mathrm{II}(A_3+3A_1)$ & --- & $W_8$ \\
        16 & $\mathrm{HET}(-16B)$ & --- & $W_{16}$ \\
        23 & $\mathrm{HET}(23A/23B)$ & --- & $W_{23}$ \\
        28 & $\mathrm{HET}(-28A)$ & --- & $W_{28},W_7,W_4$ \\
        \bottomrule
    \end{tabular}
    \caption{The remaining frame-shape orbits: two-node orbits above the dividing line and fixed nodes below it. The last two columns distinguish operations that exchange the two nodes from those that fix each node. The identity $W_1$ is omitted.}
    \label{tab:orbits-small}
\end{table}

\subsection{S-duality of charge lattices and archipelagos}
\label{sec:charge-lattice-duality}

The thresholds determine candidate dualities between symmetry descriptions. A fixed symmetry can, however,
admit several inequivalent electric charge lattices, so a further test is needed to identify the action on
individual islands. If two archipelagos are related by S-duality, their charge lattices must be paired by the
corresponding electromagnetic transformation. In particular, the two archipelagos must contain the same number
of islands at the level of the charge-lattice classification.

The charge lattices are obtained from the constructions described above, following the use of momentum
lattices to characterize CHL strings~\cite{Mikhailov:1998si}. Our test concerns the S-duality transformation
of the electric charge lattices $L_e$. We implement it using the integral kernels $K=L_e^*$, which encode the
same isometry data and are convenient for explicit lattice computations. We use the electromagnetic completion
\begin{equation}
    K=\ker_Iv,\qquad L_e=K^*,\qquad L_m=K,\qquad
    L_{\mathrm{em}}=K^*\oplus K.
\end{equation}
We assume this electromagnetic completion by analogy with the $\lambda=1$ CHL models of~\cite{Persson:2015jka}. The antisymmetric Dirac pairing is
\begin{equation}
    \langle(Q,P),(Q',P')\rangle=Q\cdot P'-Q'\cdot P.
\end{equation}
We can therefore test the dualities predicted by the frame shapes directly on the lattices, without further enumeration of shifts.

In the conventions of~\cite{Persson:2015jka}, Fricke S-duality acts as
\begin{equation}
    S\longmapsto-\frac{1}{nS},\qquad
    (Q,P)\longmapsto\left(-\frac{P}{\sqrt n},\sqrt n\,Q\right).
\end{equation}
For two islands exchanged by this operation, the required lattice relations are
\begin{equation}\label{eq:archipelago-Fricke}
    \widetilde L_e\simeq L_m(1/n),\qquad
    \widetilde L_m\simeq L_e(n),
    \qquad\text{or equivalently}\qquad
    \widetilde K\simeq K^*(n).
\end{equation}
As before, $L(c)$ denotes rescaling the bilinear form of $L$ by $c$. The isometries must match the
electromagnetic lattices with their electric and magnetic identifications.

To test the full set of proposed dualities, we also need the Atkin--Lehner operations $W_e$ for every exact divisor $e$ of $n$. The transformation
\begin{equation}
    S\longmapsto\frac{\alpha eS+\beta}{n\gamma S+\delta e},
    \qquad
    \alpha\delta e^2-\beta\gamma n=e
\end{equation}
with $\alpha,\beta,\gamma,\delta\in\mathbb Z$, acts on charges by~\cite{Persson:2015jka}
\begin{equation}
    \begin{pmatrix}Q'\\P'\end{pmatrix}
    =\frac{1}{\sqrt e}
    \begin{pmatrix}\delta e&-\beta\\-n\gamma&\alpha e\end{pmatrix}
    \begin{pmatrix}Q\\P\end{pmatrix}.
\end{equation}
The matrix has determinant one and preserves the Dirac pairing. For each island, we compare the image of its
full electromagnetic lattice with that of an island in the archipelago selected by the transformed frame
shape. This comparison must retain any correlations between electric and magnetic charges.

A balanced frame shape predicts that Fricke duality maps an archipelago back to itself. The action on
individual islands may have both fixed points and exchanged pairs. If an island is fixed, its magnetic lattice
satisfies the $n$-modularity condition
\begin{equation}
    K\simeq K^*(n),
\end{equation}
or, equivalently, $L_e\simeq L_e^*(1/n)$. Alternatively, Fricke duality can exchange two distinct islands
whose electric and magnetic charge lattices satisfy~\eqref{eq:archipelago-Fricke}. Thus closure requires a
permutation of the islands within the archipelago, with $n$-modularity characterizing its fixed points at
the charge-lattice level. The frame shape is essential because an $n$-modular lattice can also occur in two
distinct descriptions exchanged by Fricke duality. In the $-12D$ orbit, all three kernels are $12$-modular,
yet $W_{12}$ exchanges the heterotic description with the type II description based on
$D_4(a_1)+A_2$ with lift II. Their threshold frame shapes differ, so invariance of the lattice alone would
incorrectly suggest Fricke self-duality of each description. Our task is to verify that the charge lattices transform in precisely the way predicted by the threshold analysis,
both within self-dual archipelagos and between distinct ones.

\section{Explicit results}
\label{sec:explicit-results}

We first give the classification of charge lattices and their S-duality orbits for all the constructions. The
examples that follow show explicitly how the different behaviors arise: individual Fricke self-duality at
orders twenty-three, four and eight; heterotic--type II exchange and unequal shift multiplicities at order
nine; a four-description orbit at order six; and nontrivial permutations within archipelagos at order thirty.
Throughout this section we use the positive-definite representatives $I^{+}=I(-1)$ and $K^{+}=K(-1)$, where
$K=\ker_Iv$.

\subsection{General results}
\label{sec:general-results}

We collect the results for all symmetries in Tables~\ref{tab:orbits-small} and
Figure~\ref{fig:orbits-four-node}. We impose the exact-order shift condition and classify the theories by
their electric charge lattices within the family of coinvariant lattices specified above.

Table~\ref{tab:charge-lattice-classification} summarizes the full classification. Horizontal rules separate
the S-duality orbits of symmetry descriptions. Each orbit is named by one of its heterotic members, except for
the purely type II orbits $\mathrm{II}_4$ and $\mathrm{II}_8$. The letters $A,B,C,D$ label descriptions
locally within an orbit. A comma-separated list gives the shift counts for the inequivalent invariant
lattices. The kernel count identifies isometric kernels across all these shifts and invariant lattices. The
type II order is always the Spin-lift order. The labels I and II distinguish the lifts of $D_4(a_1)+A_2$ in
the threshold-frame-shape table.

\begin{longtable}{clclrr}
\caption{Charge-lattice classification with exact-order shifts. Here $n$ is the orbifold order, including the Spin lift for type II models. Comma-separated shift counts refer to distinct invariant lattices. The last column counts kernel isometry classes, equivalently electric charge-lattice classes, after identifying repetitions across shifts and invariant lattices.}\label{tab:charge-lattice-classification}\\
\toprule $n$&Orbit&Frame&Description&\shortstack{Shift classes\\per invariant lattice}&\shortstack{Kernels\\(Charge lattices)}\\
\midrule\endfirsthead
\toprule $n$&Orbit&Frame&Description&\shortstack{Shift classes\\per invariant lattice}&\shortstack{Kernels\\(Charge lattices)}\\
\midrule\endhead
4&$\mathrm{II}_4$&A&$\mathrm{II}(6A_1)$&1&1\\
\midrule
6&$-6F$&A&$\mathrm{HET}(-6F)$&2&1\\*
6&$-6F$&B&$\mathrm{II}(E_6(a_2))$&1&1\\*
6&$-6F$&C&$\mathrm{II}(3A_2)$&1&1\\*
6&$-6F$&D&$\mathrm{II}(A_2+4A_1)$&2&1\\
\midrule
8&$\mathrm{II}_8$&A&$\mathrm{II}(A_3+3A_1)$&6&4\\
\midrule
8&$8D$&A&$\mathrm{HET}(8D)$&2&1\\*
8&$8D$&B&$\mathrm{II}(D_6(a_1))$&1&1\\
\midrule
9&$9B$&A&$\mathrm{HET}(9B)$&3&1\\*
9&$9B$&B&$\mathrm{II}(E_6(a_1))$&1&1\\
\midrule
12&$-12D$&A&$\mathrm{HET}(-12D)$&8&3\\*
12&$-12D$&B&$\mathrm{II}(D_4(a_1)+A_2\text{ (II)})$&8&3\\
\midrule
12&$12G$&A&$\mathrm{HET}(12G)$&8&2\\*
12&$12G$&B&$\mathrm{II}(E_6)$&2&2\\
\midrule
12&$12H$&A&$\mathrm{HET}(12H)$&10, 16&5\\*
12&$12H$&B&$\mathrm{II}(D_6(a_2))$&8&5\\*
12&$12H$&C&$\mathrm{II}(2A_2+2A_1)$&10, 16&5\\*
12&$12H$&D&$\mathrm{II}(D_4+2A_1)$&8&5\\
\midrule
12&$-12K$&A&$\mathrm{HET}(-12K)$&8&3\\*
12&$-12K$&B&$\mathrm{II}(D_4(a_1)+A_2\text{ (I)})$&8&3\\
\midrule
14&$-14B$&A&$\mathrm{HET}(-14B)$&29&8\\*
14&$-14B$&B&$\mathrm{II}(A_6)$&11&8\\
\midrule
15&$15E$&A&$\mathrm{HET}(15E)$&18&9\\*
15&$15E$&B&$\mathrm{II}(A_4+A_2)$&18&9\\
\midrule
16&$-16B$&A&$\mathrm{HET}(-16B)$&10&2\\
\midrule
18&$-18B$&A&$\mathrm{HET}(-18B)$&21&3\\*
18&$-18B$&B&$\mathrm{HET}(18C)$&20&3\\*
18&$-18B$&C&$\mathrm{HET}(-18C)$&4, 24, 34&3\\*
18&$-18B$&D&$\mathrm{II}(E_6(a_1))$&7&3\\
\midrule
20&$20C$&A&$\mathrm{HET}(20C)$&22, 66&19\\*
20&$20C$&B&$\mathrm{II}(A_4+2A_1)$&22, 66&19\\
\midrule
20&$-20C$&A&$\mathrm{HET}(-20C)$&15, 87, 50, 26&11\\*
20&$-20C$&B&$\mathrm{II}(D_6)$&24&11\\
\midrule
23&$23A/23B$&A&$\mathrm{HET}(23A/23B)$&84&29\\
\midrule
24&$24F$&A&$\mathrm{HET}(24F)$&171, 85&35\\*
24&$24F$&B&$\mathrm{II}(A_3+A_2+A_1)$&43, 218&35\\
\midrule
24&$-24F$&A&$\mathrm{HET}(-24F)$&69, 230, 269&29\\*
24&$-24F$&B&$\mathrm{II}(D_5(a_1)+A_1)$&117&29\\
\midrule
28&$-28A$&A&$\mathrm{HET}(-28A)$&299&35\\
\midrule
30&$30D$&A&$\mathrm{HET}(30D)$&218, 807&40\\*
30&$30D$&B&$\mathrm{HET}(-30D)$&539, 486&40\\*
30&$30D$&C&$\mathrm{HET}(-30E)$&274&40\\*
30&$30D$&D&$\mathrm{II}(A_4+A_2)$&275&40\\
\bottomrule
\end{longtable}

A fixed type II symmetry can admit several inequivalent invariant lattices, each of which must be included in
the shift enumeration. For example, the $\mathrm{II}(2A_2+2A_1)$ description in the $12H$ orbit has two
invariant-lattice classes, with root systems $D_4$ and $2A_2+2A_1$, giving ten and sixteen shift classes.
Together they yield the five electric charge lattices listed in the table. This illustrates why the
classification requires all allowed invariant lattices for each symmetry.

\paragraph{Genus specification.}
See Appendix~\ref{app:lattice-genera} for genus definitions and notation.
For every symmetry description considered here, the set of positive-definite kernels $K^{+}$ is a complete
genus. Consequently the electric charge lattices $(K^{+})^*$, with the physical sign restored, also exhaust
the corresponding dual genus. Thus our enumeration gives a complete classification of the charge lattices in
these genera, including every isometry class with the specified local lattice data. The classification
applies to the constructions and shift conditions specified here.

Table~\ref{tab:remaining-genera} specifies the invariant lattices and kernels without long lists of Gram
matrices. Every entry represents the \emph{complete} genus of even Euclidean rank-six lattices with the
indicated symbol, with the common prefix $\mathrm{II}_{6,0}$ suppressed. Its invariant class number is the
length of the corresponding shift-count list, and its kernel class number is given in
Table~\ref{tab:charge-lattice-classification}. Completeness can be checked by genus masses, as illustrated in
the examples below. Different two-adic decompositions can give different-looking symbols for the same genus,
as explained in Appendix~\ref{app:lattice-genera}. In particular, the $A,C$ genera for $12H$ coincide, as do
its $B,D$ genera; the two descriptions of $20C$ also have the same kernel genus.

\begin{longtable}{llcc}
\caption{Complete invariant and charge-kernel genera. The positive-definite kernel $K^{+}$ determines the electric charge lattice by duality and reversal of sign. The prefix $\mathrm{II}_{6,0}$ is suppressed; frame labels refer to Table~\ref{tab:charge-lattice-classification}.}\label{tab:remaining-genera}\\
\toprule Orbit&Frame&\shortstack{Invariant-lattice genus\\$I^{+}$}&\shortstack{Charge-kernel genus\\$K^{+}$}\\\midrule\endfirsthead
\toprule Orbit&Frame&\shortstack{Invariant-lattice genus\\$I^{+}$}&\shortstack{Charge-kernel genus\\$K^{+}$}\\\midrule\endhead
$\mathrm{II}_4$&A&$(2_{2}^{-2})$&$(2_{6}^{+2} 4_{\mathrm{II}}^{-2})$\\
$-6F$&A&$(2_{\mathrm{II}}^{-2} 3^{-1})$&$(2_{\mathrm{II}}^{-4} 3^{+3})$\\
$-6F$&B&$(3^{+1})$&$(2_{\mathrm{II}}^{+2} 3^{-3})$\\
$-6F$&C&$(3^{+1})$&$(2_{\mathrm{II}}^{+2} 3^{-3})$\\
$-6F$&D&$(2_{\mathrm{II}}^{-2} 3^{-1})$&$(2_{\mathrm{II}}^{-4} 3^{+3})$\\
$\mathrm{II}_8$&A&$(2_{1}^{+1} 4_{5}^{-1})$&$(2_{1}^{+1} 4_{1}^{+1} 8_{\mathrm{II}}^{-2})$\\
$8D$&A&$(4_{6}^{+2})$&$(4_{6}^{+2} 8_{\mathrm{II}}^{+2})$\\
$8D$&B&$(2_{2}^{-2})$&$(2_{2}^{-2} 8_{\mathrm{II}}^{+2})$\\
$9B$&A&$(3^{+1} 9^{+1})$&$(3^{+1} 9^{-3})$\\
$9B$&B&$(3^{+1})$&$(3^{+1} 9^{-2})$\\
$-12D$&A&$(2_{\mathrm{II}}^{-2} 3^{-1})$&$(2_{\mathrm{II}}^{-2} 4_{\mathrm{II}}^{+2} 3^{+3})$\\
$-12D$&B&$(2_{\mathrm{II}}^{-2} 3^{-1})$&$(2_{\mathrm{II}}^{-2} 4_{\mathrm{II}}^{+2} 3^{+3})$\\
$12G$&A&$(4_{\mathrm{II}}^{-2} 3^{+1})$&$(4_{\mathrm{II}}^{-4} 3^{-3})$\\
$12G$&B&$(3^{+1})$&$(4_{\mathrm{II}}^{+2} 3^{-3})$\\
$12H$&A&$(2_{6}^{-2} 3^{+2})$&$(2_{6}^{-2} 4_{\mathrm{II}}^{+2} 3^{-4})$\\
$12H$&B&$(2_{2}^{-2})$&$(2_{6}^{+2} 4_{\mathrm{II}}^{+2} 3^{-2})$\\
$12H$&C&$(2_{6}^{-2} 3^{+2})$&$(2_{6}^{-2} 4_{\mathrm{II}}^{+2} 3^{-4})$\\
$12H$&D&$(2_{2}^{-2})$&$(2_{2}^{-2} 4_{\mathrm{II}}^{-2} 3^{-2})$\\
$-12K$&A&$(2_{\mathrm{II}}^{-2} 3^{-1})$&$(2_{\mathrm{II}}^{-2} 4_{\mathrm{II}}^{+2} 3^{+3})$\\
$-12K$&B&$(2_{\mathrm{II}}^{-2} 3^{-1})$&$(2_{\mathrm{II}}^{-2} 4_{\mathrm{II}}^{+2} 3^{+3})$\\
$-14B$&A&$(2_{\mathrm{II}}^{+2} 7^{-1})$&$(2_{\mathrm{II}}^{+4} 7^{+3})$\\
$-14B$&B&$(7^{-1})$&$(2_{\mathrm{II}}^{+2} 7^{+3})$\\
$15E$&A&$(3^{+1} 5^{-1})$&$(3^{-3} 5^{-3})$\\
$15E$&B&$(3^{-1} 5^{+1})$&$(3^{+3} 5^{+3})$\\
$-16B$&A&$(4_{6}^{+2})$&$(4_{6}^{+2} 16_{\mathrm{II}}^{+2})$\\
$-18B$&A&$(3^{+1} 9^{+1})$&$(2_{\mathrm{II}}^{+2} 3^{+1} 9^{-3})$\\
$-18B$&B&$(2_{\mathrm{II}}^{-2} 3^{-1})$&$(2_{\mathrm{II}}^{-4} 3^{-1} 9^{-2})$\\
$-18B$&C&$(2_{\mathrm{II}}^{-2} 3^{-1} 9^{-1})$&$(2_{\mathrm{II}}^{-4} 3^{-1} 9^{+3})$\\
$-18B$&D&$(3^{+1})$&$(2_{\mathrm{II}}^{+2} 3^{+1} 9^{-2})$\\
$20C$&A&$(2_{2}^{+2} 5^{+1})$&$(2_{6}^{-2} 4_{\mathrm{II}}^{-2} 5^{+3})$\\
$20C$&B&$(2_{2}^{+2} 5^{+1})$&$(2_{6}^{-2} 4_{\mathrm{II}}^{+2} 5^{+3})$\\
$-20C$&A&$(2_{2}^{-2} 5^{+2})$&$(2_{2}^{-2} 4_{\mathrm{II}}^{+2} 5^{+4})$\\
$-20C$&B&$(2_{2}^{-2})$&$(2_{2}^{-2} 4_{\mathrm{II}}^{+2} 5^{+2})$\\
$23A/23B$&A&$(23^{-1})$&$(23^{+3})$\\
$24F$&A&$(2_{1}^{+1} 4_{7}^{+1} 3^{+1})$&$(2_{5}^{-1} 4_{7}^{+1} 8_{\mathrm{II}}^{+2} 3^{-3})$\\
$24F$&B&$(2_{3}^{-1} 4_{5}^{-1} 3^{-1})$&$(2_{7}^{+1} 4_{5}^{-1} 8_{\mathrm{II}}^{+2} 3^{+3})$\\
$-24F$&A&$(2_{3}^{-1} 4_{7}^{+1} 3^{-2})$&$(2_{3}^{-1} 4_{7}^{+1} 8_{\mathrm{II}}^{+2} 3^{+4})$\\
$-24F$&B&$(2_{1}^{+1} 4_{5}^{-1})$&$(2_{1}^{+1} 4_{5}^{-1} 8_{\mathrm{II}}^{+2} 3^{-2})$\\
$-28A$&A&$(2_{\mathrm{II}}^{+2} 7^{-1})$&$(2_{\mathrm{II}}^{+2} 4_{\mathrm{II}}^{+2} 7^{+3})$\\
$30D$&A&$(2_{\mathrm{II}}^{+2} 3^{-1} 5^{+1})$&$(2_{\mathrm{II}}^{+4} 3^{+3} 5^{+3})$\\
$30D$&B&$(2_{\mathrm{II}}^{+2} 3^{+1} 5^{-1})$&$(2_{\mathrm{II}}^{+4} 3^{-3} 5^{-3})$\\
$30D$&C&$(3^{+1} 5^{-1})$&$(2_{\mathrm{II}}^{+2} 3^{-3} 5^{-3})$\\
$30D$&D&$(3^{-1} 5^{+1})$&$(2_{\mathrm{II}}^{+2} 3^{+3} 5^{+3})$\\
\bottomrule
\end{longtable}

\paragraph{Circle compactification and complete genera.}
\label{par:circle-genera}
The occurrence of complete genera has a natural interpretation after compactification on an additional
spectator circle. Its momentum and winding charges contribute the even unimodular hyperbolic plane
$U=\Gamma_{1,1}$. The perturbative electric lattice and its dual then take the form
\begin{equation}
 L_{e,S^1}^{\mathrm{pert}}=U\oplus K^*,\qquad
 \bigl(L_{e,S^1}^{\mathrm{pert}}\bigr)^*=U\oplus K.
\end{equation}
Here $K=K^{+}(-1)$ is the negative-definite kernel, so $U\oplus K$ has signature $(1,7)$. If $K_1$ and $K_2$
belong to the same genus, Nikulin's uniqueness theorem for even indefinite
lattices~\cite{NikulinIntegralForms} gives
\begin{equation}\label{eq:circle-genus-isometry}
 U\oplus K_1\simeq U\oplus K_2.
\end{equation}
Indeed, taking the orthogonal direct sum with $U$ preserves the discriminant form, while the resulting rank is
eight. The discriminant group $A_K$ requires at most six generators, so the sufficient uniqueness condition
$\operatorname{rank}(U\oplus K)\geq\ell(A_K)+2$ holds, where $\ell$ denotes the minimum number of generators.

A circle polarization selects a primitive copy of $U$ in this common lattice, with null generators $e,f$
satisfying $e\cdot f=1$. Its orthogonal complement is the four-dimensional kernel recovered in the
corresponding decompactification limit. Equivalently, selecting the primitive momentum direction $e$, of
divisibility one, gives
\begin{equation}
 e^\perp/\mathbb Z e\simeq K_i.
\end{equation}
Divisibility one means that the pairings of $e$ with the lattice generate $\mathbb Z$, which permits the
unimodular momentum--winding splitting. Equation~\eqref{eq:circle-genus-isometry} realizes every member of the
genus as the complement of such a polarization of the same lattice. Conversely, every such complement has the
same signature and discriminant form and hence lies in that genus. Nonisometric complements give inequivalent
polarizations. Completeness therefore means that our list exhausts the kernels of these ordinary circle
decompactification limits at the charge-lattice level. Interpreting them as limits of a single
three-dimensional theory additionally requires an identification of the full compactified theories. Related
analyses of shift orbifolds and decompactification limits in Narain theories appear in~\cite{Israel:2025ouq}.
For CHL models, a complementary three-dimensional perspective is provided by U-duality constraints on
four-derivative scalar couplings~\cite{Bossard:2017wum}.

\paragraph{The action on individual islands.}
The frame shapes identify which archipelagos can be related by duality. We now determine how the individual
islands are paired. Applying the charge transformation of Section~\ref{sec:charge-lattice-duality} to an
island with magnetic lattice $K$ gives the magnetic lattice of its proposed $W_e$-dual. For every exact
divisor $e\mid n$, its positive-definite representative is
\begin{equation}\label{eq:remaining-partial-dual}
 \mathcal D_e(K^{+})=\left(K^{+}+\frac{n}{e}(K^{+})^*\right)(e).
\end{equation}
The sum is taken inside $K^{+}\otimes\mathbb Q$, and the final $(e)$ rescales the bilinear form by $e$.
For $e=n$, this reduces to the Fricke transformation $(K^{+})^*(n)$. For the other exact divisors, it
specifies the corresponding Atkin--Lehner pairing. Integral isometry tests verify that the transformed
lattices belong to the predicted target genus and exhaust its classes.

An operation that preserves the frame shape can still exchange distinct islands within the same
archipelago. To describe these permutations for a two-description orbit, choose an exchanging divisor
$e_0$ and label the paired kernels by $B_i=\mathcal D_{e_0}(A_i)$. The other independent operation $W_f$
preserves the description and induces an involution $\sigma$ on the indices:
\begin{equation}
 W_{e_0}:A_i\leftrightarrow B_i,\qquad
 W_f:A_i\mapsto A_{\sigma(i)},\quad B_i\mapsto B_{\sigma(i)}.
\end{equation}
The product acts by exchanging descriptions and applying $\sigma$. Table~\ref{tab:remaining-actions} records
its cycle type, with $1^a2^b$ denoting $a$ fixed kernels and $b$ exchanged pairs. Thus each fixed index gives a
two-member orbit, and each transposition a four-member orbit. Formula~\eqref{eq:remaining-partial-dual},
together with the genus specification, determines which actual lattices are paired without relying on an
arbitrary numbering of the classes.

\begin{center}
\begin{tabular}{lccc}
\toprule
Orbit&\shortstack{Exchanging divisor\\$e_0$}&\shortstack{Preserving divisor\\$f$}&\shortstack{Kernel permutation\\$\sigma$}\\
\midrule
$8D$&8&---&$1^{1}$\\
$9B$&9&---&$1^{1}$\\
$-12D$&3&4&$1^{1}2^{1}$\\
$12G$&4&3&$1^{2}$\\
$-12K$&3&12&$1^{3}$\\
$-14B$&2&7&$1^{8}$\\
$15E$&3&15&$1^{9}$\\
$20C$&4&20&$1^{13}2^{3}$\\
$-20C$&5&4&$1^{5}2^{3}$\\
$24F$&3&24&$1^{19}2^{8}$\\
$-24F$&3&8&$1^{15}2^{7}$\\
\bottomrule
\end{tabular}
\captionsetup{hypcap=false}
\captionof{table}{Generators and kernel permutations for the two-description orbits. The operation $W_{e_0}$ exchanges the descriptions, while $W_f$ preserves each description and permutes its charge lattices. The notation $1^a2^b$ denotes $a$ fixed lattices and $b$ exchanged pairs.}
\label{tab:remaining-actions}
\end{center}

For the four-description $12H$ orbit choose
\begin{equation}
 B_i=\mathcal D_{12}(A_i),\qquad C_i=\mathcal D_4(A_i),\qquad
 D_i=\mathcal D_3(A_i),\qquad i=1,\ldots,5.
\end{equation}
Then $W_4$ exchanges $A_i,C_i$ and $B_i,D_i$, while $W_3$ exchanges $A_i,D_i$ and $B_i,C_i$. No index
permutation is needed in this convention. This convention is possible because the two type II descriptions
with coinvariant lattice $D_6$ have the same set of kernels. There are five four-member orbits.

The single-description $-28A$ case has a different pattern. Of its $35$ kernels, nine are fixed by every
Atkin--Lehner operation. Eleven pairs are exchanged by $W_4$ and $W_{28}$, and fixed separately by $W_7$. The
remaining two pairs are exchanged by $W_4$ and $W_7$, but each member is fixed by $W_{28}$. Consequently there
are thirteen $28$-modular lattices, although only nine kernels are individually fixed by the whole
Atkin--Lehner group. The full action gives nine singleton orbits and thirteen two-member orbits. Fricke
self-duality and invariance under every partial duality are therefore distinct properties.

At orders twenty and twenty-four the balanced descriptions also contain exchanged Fricke pairs. For $20C$,
thirteen of the nineteen kernels in each description are $20$-modular and the other six form three pairs. For
$24F$, nineteen of the thirty-five kernels in each description are $24$-modular and the other sixteen form
eight pairs. The $-20C$ and $-24F$ orbits instead exhibit nontrivial permutations under the
description-preserving operations $W_4$ and $W_8$, respectively. The order-thirty example below exhibits this
mixing explicitly.

The two order-sixteen $-16B$ islands are individually Fricke self-dual. The order-eighteen orbit has the same
four-description pattern as $12H$, with three islands in each description. Using the labels in
Table~\ref{tab:charge-lattice-classification}, choose $B_i=\mathcal D_{18}(A_i)$, $C_i=\mathcal D_2(A_i)$ and
$D_i=\mathcal D_9(A_i)$. Then $W_2$ exchanges $A_i,C_i$ and $B_i,D_i$, while $W_9$ exchanges $A_i,D_i$ and
$B_i,C_i$, giving three four-member orbits.

The kernels fixed by Fricke duality at the lattice level yield a collection of even Euclidean modular
lattices, satisfying $K^{+}\simeq(K^{+})^*(n)$. To place this mathematical output of the
S-duality analysis in context, Appendix~\ref{app:modular-catalogue} compares all 121 distinct modular lattices
obtained here with the Nebe--Sloane catalogue~\cite{NebeSloaneCatalogue}, recording the matches and the scope
of the comparison. This count includes the three $12$-modular lattices shared by the two descriptions in
the $-12D$ orbit, where Fricke duality exchanges the descriptions despite preserving each lattice isometry
class. These same three lattices also occur in the $-12K$ constructions and are counted only once.

\subsection{The self-dual order-twenty-three archipelago}
\label{sec:order23}

As a first example of individual Fricke self-duality, we consider the heterotic symmetry of frame shape $1\,23$. Its frame shape is balanced. The associated Leech fixed lattice has Gram matrix $\left(\begin{smallmatrix}4&1\\1&6\end{smallmatrix}\right)$. We label the positive-definite kernels by their defining shifts, $K^{+}_v=(\ker_Iv)(-1)$.

\paragraph{Invariant lattices and shifts.}
There is a single isometry class of invariant lattices. In a basis of $I^{+}$, its Gram matrix is
\begin{equation}\label{eq:order23-invariant}
 G_{I^{+}}=\begin{pmatrix}
 2&-1&-1&-1&0&1\\
 -1&2&1&1&0&-1\\
 -1&1&2&1&0&0\\
 -1&1&1&2&0&-1\\
 0&0&0&0&2&-1\\
 1&-1&0&-1&-1&4
 \end{pmatrix},\qquad
 \det I^{+}=23,\qquad |O(I^{+})|=480.
\end{equation}
Its roots span $A_4\oplus A_1$. The uniqueness statement is at the level of invariant-lattice isometry classes.

The allowed shifts fall into $84$ equivalence classes under lattice automorphisms, translations by
$(I^{+})^*$, and changes of the orbifold generator. We explain their enumeration in
Appendix~\ref{app:shift-enumeration}.

For each representative, the positive-definite kernel is specified explicitly by
\begin{equation}\label{eq:order23-kernel}
 K^{+}_w=\{x\in\mathbb Z^6:xG_{I^{+}}w^{\mathsf T}\equiv0\pmod{23}\},
 \qquad (x,y)=xG_{I^{+}}y^{\mathsf T}.
\end{equation}
The $84$ shifts yield $29$ pairwise non-isometric kernels, denoted $K^{+}_1,\ldots,K^{+}_{29}$. Each has
determinant $23^3$ and discriminant group $(\mathbb Z/23\mathbb Z)^3$. Table~\ref{tab:order23-kernels} gives
one defining shift for each kernel, together with its multiplicity among the shift classes, its minimum
squared norm, and its automorphism-group order. Equations~\eqref{eq:order23-invariant}
and~\eqref{eq:order23-kernel} specify each integral lattice explicitly.

\paragraph{Same charge lattice implies same 1-loop partition function.}
This island is special in that it is the only case where the orbifold order is prime. Moreover, gauging
$\Theta$ without a shift recovers the parent toroidal CFT. These facts allow us to write the partition function entirely in terms of
$L_e=K_v^*$ and universal oscillator factors (see Appendix~\ref{app:order23-partition} for details).\footnote{A similar rearrangement for the CHL string was used
in Appendix~A of~\cite{Bossard:2017wum}.} Define
\begin{equation}\label{eq:order23-theta}
 \Theta_M(\tau,\bar\tau)=\sum_{Q\in M}q^{Q_L^2/2}\bar q^{Q_R^2/2},
 \qquad q=e^{2\pi i\tau},\qquad \bar q=e^{-2\pi i\bar\tau}.
\end{equation}
Here $M$ is negative definite, so $Q_L=0$ and $Q^2=-Q_R^2$. With the common right-moving and
noncompact contribution written as
\begin{equation}\label{eq:order23-universal}
 \mathcal U(\tau,\bar\tau)=\frac{\bar V_8-\bar S_8}{\tau_2\bar\eta(\bar\tau)^8},
 \qquad Z_v=\mathcal U\,\mathcal Z_v,
\end{equation}
the remaining factor is
\begin{equation}\label{eq:order23-partition}
 \boxed{\displaystyle
 \mathcal Z_v(\tau,\bar\tau)=
 \frac{\Theta_{L_e^*}(\tau,\bar\tau)}{\eta(\tau)\eta(23\tau)}
 +\frac1{23}\sum_{b=0}^{22}
 \frac{\Theta_{L_e}(\tau+b,\bar\tau+b)}
 {\eta(\tau+b)\eta\bigl((\tau+b)/23\bigr)}.}
\end{equation}
All shift dependence is through the isometry class of the charge lattice. Thus inequivalent shifts giving
isometric kernels have the same one-loop function $\mathcal Z_v$, before the universal supersymmetric factor
sets the vacuum amplitude to zero. This equality of unrefined spectra supports the conjecture that these
shift multiplicities are spurious. Equality of operator products remains to be established.

\paragraph{Kernels and genera.}
It can also be checked that these twenty-nine lattices form a complete genus, for example by comparing their
total mass with the genus mass using SageMath. The definitions of genus, class number and genus mass, together
with the local-symbol conventions, are given in Appendix~\ref{app:lattice-genera}.

Completeness allows us to specify the entire set as the isometry classes of even Euclidean lattices of rank six with genus symbol
\begin{equation}\label{eq:order23-genus}
 \boxed{\mathrm{II}_{6,0}\bigl(23^{+3}\bigr).}
\end{equation}
We use the standard convention in which $\mathrm{II}$ denotes an even lattice, the subscript records its
signature, and the remaining symbols give its non-unimodular local Jordan components. For an odd prime $p$, a
factor $(p^a)^{\epsilon r}$ denotes a rank-$r$ component whose form is $p^a$ times a unimodular form, with
determinant a quadratic residue for $\epsilon=+$ and a nonresidue for $\epsilon=-$. Unimodular components are
implicit. Here the full local symbols are $1_{\mathrm{II}}^{+6}$ at two and $1^{+3}23^{+3}$ at twenty-three.
The genus symbol specifies the full collection of twenty-nine classes. This is particularly useful for large
archipelagos, for which listing every Gram matrix would be cumbersome.

\paragraph{S-duality action.}
At the charge-lattice level, $W_{23}$ sends $K^{+}$ to $(K^{+})^*(23)$. Integral isometry tests give
\begin{equation}
 K^{+}_i\simeq (K^{+}_i)^*(23),\qquad i=1,\ldots,29.
\end{equation}
For each kernel Gram matrix $G_i$, this is verified by an integral unimodular matrix $U_i$ satisfying
$U_i(23G_i^{-1})U_i^{\mathsf T}=G_i$. Thus all twenty-nine islands are individually fixed at the
charge-lattice level, giving $29$ modular classes and no exchanged pairs.

\paragraph{Comparison with the lattice catalogue.}
The construction thus produces a complete genus of twenty-nine even Euclidean lattices of rank six, each of
which is $23$-modular. To our knowledge, this complete collection has not previously been tabulated in the
literature. We checked the six-dimensional listings, the modular-lattice tables, and the downloadable standard
records of the Nebe--Sloane catalogue~\cite{NebeSloaneCatalogue}. None of our even kernels is listed there.
The catalogue does contain a rank-six $23$-modular lattice of determinant $23^3$, named \texttt{6QF.6.e}, but
it is odd, with minimum seven, and hence is not isometric to any of our kernels. Some members of our genus are
direct sums of familiar binary lattices; the potentially new result is the complete collection and its
realization by these orbifolds.

\subsection{Self-dual type II islands of orders four and eight}
\label{sec:typeII-selfdual}

The type II symmetries $6A_1$ and $A_3+3A_1$, with Spin-lift orders four and eight, have balanced threshold frame shapes
\begin{equation}
 \pi_4=1^{-4}2^{10}4^{-4},\qquad
 \pi_8=1^{-2}2^3 4^3 8^{-2}.
\end{equation}
Each archipelago is therefore mapped to itself by Fricke duality. We now determine whether this operation fixes its individual islands or exchanges them.

\paragraph{Invariant lattices and shifts.}
The coinvariant lattices are $D_6$ and $D_5\oplus A_1$, respectively. Each construction admits a unique
invariant-lattice isometry class, with positive-definite representatives $I_4^{+}=D_6$ and $I_8^{+}=D_5\oplus
A_1$. We use the bases
\begin{equation}\label{eq:typeII48-invariants}
 G_{I_4^{+}}=\begin{pmatrix}
 2&-1&0&0&0&0\\
 -1&2&-1&0&0&0\\
 0&-1&2&-1&0&0\\
 0&0&-1&2&-1&-1\\
 0&0&0&-1&2&0\\
 0&0&0&-1&0&2
 \end{pmatrix},\qquad
 G_{I_8^{+}}=\begin{pmatrix}
 2&-1&0&0&0&0\\
 -1&2&-1&0&0&0\\
 0&-1&2&-1&-1&0\\
 0&0&-1&2&0&0\\
 0&0&-1&0&2&0\\
 0&0&0&0&0&2
 \end{pmatrix}.
\end{equation}
Their determinants are four and eight, and their automorphism-group orders are $46080$ and $7680$. Shift
enumeration gives one class at order four and six at order eight. The following table lists all
representatives $v=w/n$, together with their kernel identifications, minimum squared norms, and
automorphism-group orders:
\begin{center}
\begin{tabular}{clcrr}
\toprule
$n$&\shortstack{Shift numerator\\$w$}&\shortstack{Charge kernel\\$K^{+}$}&\shortstack{Minimum\\squared norm}&\shortstack{Automorphism\\order}\\
\midrule
4&$(2,1,0,1,0,0)$&$K^{+}_4$&2&2304\\
\midrule
8&$(3,1,0,1,0,0)$&$K^{+}_{8,1}$&2&288\\
8&$(3,1,0,0,0,1)$&$K^{+}_{8,2}$&2&256\\
8&$(2,5,1,0,0,1)$&$K^{+}_{8,3}$&4&384\\
8&$(2,0,2,1,0,1)$&$K^{+}_{8,1}$&2&288\\
8&$(2,1,0,1,0,2)$&$K^{+}_{8,4}$&2&2304\\
8&$(5,2,0,1,0,2)$&$K^{+}_{8,2}$&2&256\\
\bottomrule
\end{tabular}
\end{center}
\paragraph{Kernels and genera.}
Each kernel is defined by $xG_{I^{+}_n}w^{\mathsf T}=0\pmod n$, with the pairing inherited from $I^{+}_n$.
These data therefore specify the integral lattices without listing five further Gram matrices. There is one
island at order four and four at order eight, with kernel determinants $4^3$ and $8^3$, respectively.

The kernels exhaust the genera
\begin{equation}
 \mathcal G_4=\mathrm{II}_{6,0}(2_6^{+2}4_{\mathrm{II}}^{-2}),\qquad
 \mathcal G_8=\mathrm{II}_{6,0}(2_1^{+1}4_1^{+1}8_{\mathrm{II}}^{-2}),
\end{equation}
of class numbers one and four. Here a numerical subscript on a two-adic component denotes its \emph{oddity},
the sum, modulo eight, of the odd diagonal units of its unimodular form. The subscript $\mathrm{II}$ continues
to denote an even component. These are choices of local Jordan symbols. Equivalent decompositions can give
different-looking symbols. Genus completeness can be checked as in the preceding examples.

\paragraph{S-duality action.}
Integral isometry tests establish
\begin{equation}
 K^{+}_4\simeq (K^{+}_4)^*(4),\qquad
 K^{+}_{8,i}\simeq (K^{+}_{8,i})^*(8),\quad i=1,\ldots,4.
\end{equation}
Thus all five islands are individually Fricke self-dual at the charge-lattice level. There are no exchanged pairs in either archipelago.

Every Atkin--Lehner orbit in the CHL family considered by Persson and Volpato~\cite{Persson:2015jka} contains
a heterotic corner. The two type II orbits here do not. Since both orders are non-square-free, additional
cusps outside the Atkin--Lehner orbit remain possible locations for heterotic descriptions. We return to this
distinction and its CHL counterpart in Section~\ref{sec:discussion-results}.

\subsection{The heterotic 9B island and its type II dual}
\label{sec:order9}

We next consider the heterotic class $9B$, with frame shape $3^{-1}9^3$. Fricke duality transforms this frame
shape into $1^3 3^{-1}$, the threshold frame shape of the order-nine Spin lift of the type II symmetry
$E_6(a_1)$. The threshold analysis therefore predicts an exchange of these two archipelagos. We now check this
prediction on their charge lattices, again using the positive-definite conventions $I^{+}=I(-1)$ and
$K^{+}=(\ker_Iv)(-1)$.

\paragraph{Invariant lattices and shifts.}
The Leech fixed lattice for $9B$ has Gram matrix $\left(\begin{smallmatrix}6&3\\3&6\end{smallmatrix}\right)$. Its complement in $E_8$ gives a single invariant-lattice isometry class, represented by
\begin{equation}\label{eq:order9-invariants}
 G_{I^{+,\mathrm{het}}}=\begin{pmatrix}
 2&-1&-1&1&1&1\\
 -1&2&1&-1&0&0\\
 -1&1&2&0&-1&0\\
 1&-1&0&2&0&0\\
 1&0&-1&0&4&2\\
 1&0&0&0&2&4
 \end{pmatrix},\qquad
 G_{I^{+,\mathrm{II}}}=\begin{pmatrix}
 2&-1&0&0&0&0\\
 -1&2&-1&0&0&0\\
 0&-1&2&-1&0&-1\\
 0&0&-1&2&-1&0\\
 0&0&0&-1&2&0\\
 0&0&-1&0&0&2
 \end{pmatrix}.
\end{equation}
For the type II construction, the coinvariant lattice is $N=E_6$, and the corresponding invariant lattice is
also $I^{+,\mathrm{II}}\simeq E_6$, with the Gram matrix displayed above. The superscripts $\mathrm{het}$ and
$\mathrm{II}$ distinguish the heterotic and type II constructions. The two invariant lattices have
determinants $27$ and $3$, and automorphism-group orders $2304$ and $103680$, respectively.

Applying the enumeration of Appendix~\ref{app:shift-enumeration} gives three heterotic shift classes and one
type II shift class. Representatives $v=w/9$, in the respective bases of~\eqref{eq:order9-invariants}, are
\begin{equation}\label{eq:order9-shifts}
 \begin{split}
 w_{\mathrm{het},1}&=(5,1,1,0,0,0),\qquad
 w_{\mathrm{het},2}=(4,2,0,0,1,0),\\
 w_{\mathrm{het},3}&=(2,1,0,1,1,0),\qquad
 w_{\mathrm{II}}=(5,2,1,0,0,0).
 \end{split}
\end{equation}
All satisfy level matching and the exact-order condition on the shift character. Although the three heterotic
shifts are inequivalent under the identifications used here, their kernels are integrally isometric. Thus
there is precisely one island on each side at the charge-lattice level.

The fact that there are three inequivalent shifts on the heterotic side but only one on the type II side
suggests, through S-duality, that the three heterotic shifts in fact give rise to the same CFT. Establishing
this proposed CFT equivalence would require a comparison beyond the charge lattice.

\paragraph{Kernels and genera.}
Gram matrices for the two kernels in convenient integral bases are
\begin{equation}\label{eq:order9-kernels}
 G_{K^{+,\mathrm{het}}}=\begin{pmatrix}
 2&0&0&-1&-1&-1\\
 0&4&-2&-2&-1&1\\
 0&-2&4&1&-1&-2\\
 -1&-2&1&6&1&0\\
 -1&-1&-1&1&6&1\\
 -1&1&-2&0&1&6
 \end{pmatrix},\qquad
 G_{K^{+,\mathrm{II}}}=\begin{pmatrix}
 2&0&0&-1&0&-1\\
 0&2&0&-1&-1&-1\\
 0&0&2&0&-1&-1\\
 -1&-1&0&4&-1&1\\
 0&-1&-1&-1&4&1\\
 -1&-1&-1&1&1&6
 \end{pmatrix}.
\end{equation}
Their determinants are $3^7$ and $3^5$, respectively, as required for index-nine sublattices of the
corresponding invariant lattices. Both have automorphism-group order $192$. Each kernel is the unique class in
its own genus. This can also be checked by a genus-mass calculation, for example using SageMath.
In the notation of~\eqref{eq:order23-genus}, these genera are
\begin{equation}
 \operatorname{gen}(K^{+,\mathrm{het}})=\mathrm{II}_{6,0}(3^{+1}9^{-3}),
 \qquad
 \operatorname{gen}(K^{+,\mathrm{II}})=\mathrm{II}_{6,0}(3^{+1}9^{-2}).
\end{equation}

\paragraph{S-duality action.}
An explicit integral isometry, given in Appendix~\ref{app:shift-enumeration}, establishes
\begin{equation}\label{eq:order9-Fricke}
 \boxed{K^{+,\mathrm{II}}\simeq (K^{+,\mathrm{het}})^*(9),
 \qquad K^{+,\mathrm{het}}\simeq (K^{+,\mathrm{II}})^*(9).}
\end{equation}
Consequently, the electric and magnetic lattices transform precisely as in~\eqref{eq:archipelago-Fricke}. The
determinant relation $9^6/3^7=3^5$ provides a further check of the integral isometry. Neither kernel is
$9$-modular, since a rank-six $9$-modular lattice would have determinant $9^3=3^6$. Instead, the two distinct
kernels form one exchanged pair, in agreement with the two frame shapes.

In particular, the type II model is the order-nine string island constructed by Dabholkar and
Harvey~\cite{Dabholkar:1998kv}, based on $\Gamma_{6,6}(E_6)$ and the class $E_6(a_1)$. Our analysis identifies
its heterotic S-dual with the $9B$ island and verifies the predicted relation between their charge lattices.

\subsection{The order-six orbit: \texorpdfstring{$-6F$}{-6F} and three type II descriptions}
\label{sec:order6}

This example shows why charge lattices alone do not determine the duality action. Four distinct symmetry
descriptions share only two charge-lattice isometry classes. The threshold frame shapes distinguish the
candidate frames and specify which one is reached by each Atkin--Lehner operation. The lattice comparison
then tests these assignments.

The heterotic class $-6F$ has frame shape $1\,2^{-2}3^{-3}6^6$. Its three nontrivial Atkin--Lehner images are
the order-six type II models with Carter classes $3A_2$, $A_2+4A_1$, and $E_6(a_2)$, reached by $W_2$, $W_3$,
and $W_6$, respectively.

\begin{figure}[htbp]
\centering
\begin{tikzpicture}[x=11cm,y=11cm,font=\small]
\draw[->,thin] (-.06,0)--(1.12,0) node[right] {$\operatorname{Re}S$};
\draw[line width=.8pt] (0,0)--(0,.49) (1,0)--(1,.49);
\draw[dashed,line width=.8pt] (0,.49)--(0,.61) (1,.49)--(1,.61);
\node[anchor=east] at (-.015,.53) {$\operatorname{Im}S$};
\node[anchor=east] at (-.012,.36) {$a$};
\node[anchor=west] at (1.012,.36) {$a^{\prime}$};
\node at (.5,.36) {$\mathcal F_{6}$};
\draw[->,thin] (.5,.48)--(.5,.59);
\node[align=center,anchor=south] at (.5,.605) {$i\infty$\\[3pt]$\mathrm{HET}(-6F)$};
\draw[line width=.8pt] (0.00000000,0) arc[start angle=180,end angle=0,radius=0.16666667];
\node[inner sep=1pt,anchor=south] at (0.16666667,0.17866667) {$b$};
\draw[line width=.8pt] (0.33333333,0) arc[start angle=180,end angle=0,radius=0.08333333];
\node[inner sep=1pt,anchor=south] at (0.41666667,0.09533333) {$c$};
\draw[line width=.8pt] (0.50000000,0) arc[start angle=180,end angle=0,radius=0.08333333];
\node[inner sep=1pt,anchor=south] at (0.58333333,0.09533333) {$c^{\prime}$};
\draw[line width=.8pt] (0.66666667,0) arc[start angle=180,end angle=0,radius=0.16666667];
\node[inner sep=1pt,anchor=south] at (0.83333333,0.17866667) {$b^{\prime}$};
\draw[fill=white,line width=.6pt] (0.00000000,0) circle[radius=.0035];
\node[anchor=north,inner sep=3pt] at (0.00000000,-.006) {$0$};
\draw[fill=white,line width=.6pt] (0.33333333,0) circle[radius=.0035];
\node[anchor=north,inner sep=3pt] at (0.33333333,-.006) {$\frac{1}{3}$};
\draw[fill=white,line width=.6pt] (0.50000000,0) circle[radius=.0035];
\node[anchor=north,inner sep=3pt] at (0.50000000,-.006) {$\frac{1}{2}$};
\draw[fill=white,line width=.6pt] (0.66666667,0) circle[radius=.0035];
\node[anchor=north,inner sep=3pt] at (0.66666667,-.006) {$\frac{2}{3}$};
\draw[fill=white,line width=.6pt] (1.00000000,0) circle[radius=.0035];
\node[anchor=north,inner sep=3pt] at (1.00000000,-.006) {$1$};
\node[align=center,anchor=north,font=\footnotesize] at (0.00000000,-0.085) {$\mathrm{II}(E_6(a_2))$\\[2pt]$W_6$};
\node[align=center,anchor=north,font=\footnotesize] at (0.33333333,-0.085) {$\mathrm{II}(3A_2)$\\[2pt]$W_2$};
\node[align=center,anchor=north,font=\footnotesize] at (0.50000000,-0.145) {$\mathrm{II}(A_2+4A_1)$\\[2pt]$W_3$};
\node[align=center,anchor=north,font=\footnotesize] at (0.66666667,-0.085) {$\sim\frac13$};
\node[align=center,anchor=north,font=\footnotesize] at (1.00000000,-0.085) {$\sim0$};
\end{tikzpicture}
\caption{A fundamental domain $\mathcal F_6$ for $\Gamma_0(6)$ in the axiodilaton upper half-plane. The region lies between the two vertical sides and above the four semicircles. The vertical sides extend to infinity. Sides with matching letters are identified. The vertices $0\sim1$ and $1/3\sim2/3$ give four inequivalent cusps in total. Their proposed string descriptions are labeled, with $W_6,W_2,W_3$ mapping the heterotic cusp at infinity to $0,1/3,1/2$, respectively.}
\label{fig:order6-axiodilaton}
\end{figure}
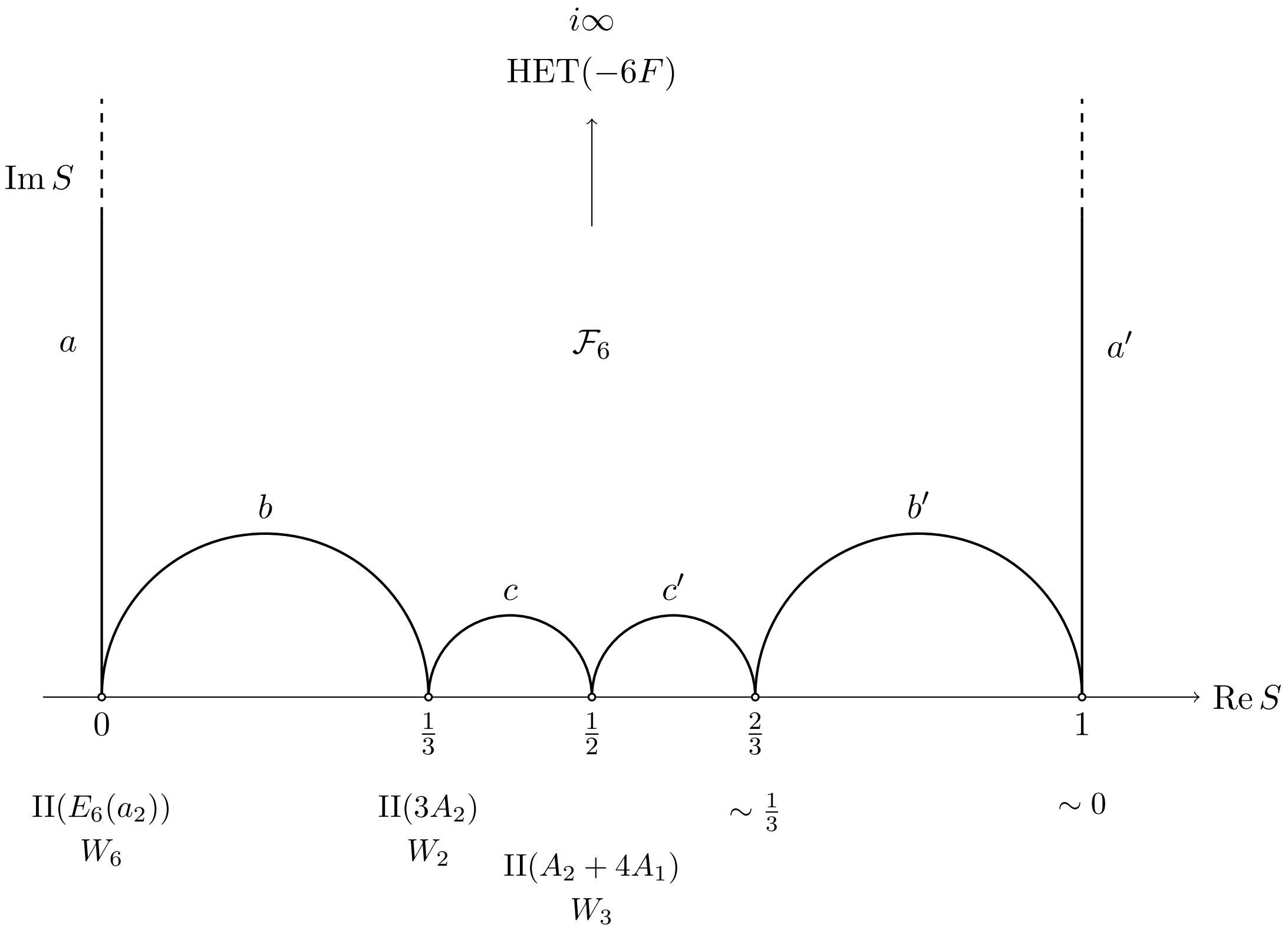

Figure~\ref{fig:order6-axiodilaton} displays the four cusp limits of the axiodilaton in the heterotic
coordinate $S$. The heterotic description occurs at infinity, and $W_2$, $W_3$ and $W_6$ map it to the type II
descriptions at $1/3$, $1/2$ and $0$, respectively.

\paragraph{Invariant lattices and shifts.}
For $-6F$, the Leech fixed lattice has Gram matrix $\left(\begin{smallmatrix}4&2\\2&4\end{smallmatrix}\right)$. Its complement gives the unique invariant-lattice class $I_D^{+}\simeq D_4\oplus A_2$, with
\begin{equation}\label{eq:order6-invariant}
 G_{I_D^{+}}=\begin{pmatrix}
 2&-1&-1&1&0&0\\
 -1&2&1&-1&0&0\\
 -1&1&2&0&0&0\\
 1&-1&0&2&0&0\\
 0&0&0&0&2&-1\\
 0&0&0&0&-1&2
 \end{pmatrix},\qquad
 \det I_D^{+}=12,\qquad |O(I_D^{+})|=13824.
\end{equation}
The type II class $A_2+4A_1$ has $N=D_4\oplus A_2$ and the same invariant lattice $I_D^{+}$. The other two
type II classes have $N=E_6$, with invariant lattice $I_E^{+}\simeq E_6$. We use the $E_6$ basis
in~\eqref{eq:order9-invariants}. In each case there is a single invariant-lattice isometry class.

The allowed shifts $v=w/6$ on $I_D^{+}$ have two inequivalent representatives,
\begin{equation}
 u_1=(4,1,1,0,1,0),\qquad u_2=(2,1,0,0,3,0),
\end{equation}
whose kernels are isometric. On $I_E^{+}$, there is one shift class, represented by
\begin{equation}
 t=(3,2,1,0,0,0).
\end{equation}
The enumeration therefore gives one kernel class for each of the four symmetry descriptions:
\begin{center}
\begin{tabular}{lccc}
\toprule
Symmetry & Invariant lattice & Shift numerators & Kernel\\
\midrule
$\mathrm{HET}(-6F)$ & $I_D^{+}$ & $u_1,u_2$ & $K^{+}_D$\\
$\mathrm{II}(3A_2)$ & $I_E^{+}$ & $t$ & $K^{+}_E$\\
$\mathrm{II}(A_2+4A_1)$ & $I_D^{+}$ & $u_1,u_2$ & $K^{+}_D$\\
$\mathrm{II}(E_6(a_2))$ & $I_E^{+}$ & $t$ & $K^{+}_E$\\
\bottomrule
\end{tabular}
\end{center}

\paragraph{Kernels and genera.}
In convenient integral bases, the two abstract kernels have Gram matrices
\begin{equation}\label{eq:order6-kernels}
 G_{K^{+}_D}=\begin{pmatrix}
 2&0&0&0&-1&-1\\
 0&2&0&0&-1&-1\\
 0&0&4&-2&0&0\\
 0&0&-2&4&0&0\\
 -1&-1&0&0&4&1\\
 -1&-1&0&0&1&4
 \end{pmatrix},\qquad
 G_{K^{+}_E}=\begin{pmatrix}
 2&-1&0&0&0&0\\
 -1&2&0&0&0&0\\
 0&0&2&0&-1&-1\\
 0&0&0&2&-1&-1\\
 0&0&-1&-1&4&1\\
 0&0&-1&-1&1&4
 \end{pmatrix}.
\end{equation}
Their determinants are $432=2^4 3^3$ and $108=2^2 3^3$, respectively. Both automorphism groups have order
$768$. Each kernel is the unique class in its genus, so they can equivalently be specified by
\begin{equation}\label{eq:order6-genera}
 \operatorname{gen}(K^{+}_D)=\mathrm{II}_{6,0}(2_{\mathrm{II}}^{-4}3^{+3}),
 \qquad
 \operatorname{gen}(K^{+}_E)=\mathrm{II}_{6,0}(2_{\mathrm{II}}^{+2}3^{-3}).
\end{equation}
For the two-adic factors, the subscript $\mathrm{II}$ means that the underlying unimodular component is even. Its sign is positive when its determinant is $\pm1\pmod8$ and negative when it is $\pm3\pmod8$. As in the
preceding examples, genus completeness can be checked by a genus-mass calculation.

\paragraph{S-duality action.}
To test the partial operations as well as Fricke duality, define, for $e=2,3,6$,
\begin{equation}\label{eq:order6-partial-dual}
 \mathcal D_e(K^{+})=\left(K^{+}+\frac{6}{e}(K^{+})^*\right)(e)
 =\left((K^{+})^*\cap\frac1e K^{+}\right)(e).
\end{equation}
Here both kernels satisfy $6(K^{+})^*\subseteq K^{+}$. Since $e$ and $6/e$ are coprime, the two descriptions
agree, and the charge transformation sends the full electromagnetic lattice to
\begin{equation}
 (K^{+})^*\oplus K^{+}\ \longmapsto\ \mathcal D_e(K^{+})^*\oplus\mathcal D_e(K^{+}).
\end{equation}
In particular, $\mathcal D_6(K^{+})=(K^{+})^*(6)$. Integral isometry tests give
\begin{equation}\label{eq:order6-dualities}
 \begin{aligned}
 \mathcal D_2(K^{+}_D)&\simeq K^{+}_E,&
 \mathcal D_3(K^{+}_D)&\simeq K^{+}_D,&
 \mathcal D_6(K^{+}_D)&\simeq K^{+}_E,\\
 \mathcal D_2(K^{+}_E)&\simeq K^{+}_D,&
 \mathcal D_3(K^{+}_E)&\simeq K^{+}_E,&
 \mathcal D_6(K^{+}_E)&\simeq K^{+}_D.
 \end{aligned}
\end{equation}
Thus every edge of the frame-shape orbit has the required charge-lattice pairing:
\begin{center}
\small
\begin{tabular}{lccc}
\toprule
Starting description & $W_2$ image & $W_3$ image & $W_6$ image\\
\midrule
$\mathrm{HET}(-6F)$ & $\mathrm{II}(3A_2)$ & $\mathrm{II}(A_2+4A_1)$ & $\mathrm{II}(E_6(a_2))$\\
$\mathrm{II}(3A_2)$ & $\mathrm{HET}(-6F)$ & $\mathrm{II}(E_6(a_2))$ & $\mathrm{II}(A_2+4A_1)$\\
$\mathrm{II}(A_2+4A_1)$ & $\mathrm{II}(E_6(a_2))$ & $\mathrm{HET}(-6F)$ & $\mathrm{II}(3A_2)$\\
$\mathrm{II}(E_6(a_2))$ & $\mathrm{II}(A_2+4A_1)$ & $\mathrm{II}(3A_2)$ & $\mathrm{HET}(-6F)$\\
\bottomrule
\end{tabular}
\end{center}
In particular, $W_3$ preserves each abstract kernel while exchanging distinct symmetry descriptions. Thus
distinct frame shapes can share the same charge lattice. Fricke duality exchanges $K^{+}_D$ and $K^{+}_E$;
neither is $6$-modular, since neither has determinant $6^3$. There is one island in each archipelago at the
charge-lattice level, with all three dualities closing as predicted.

\subsection{The order-thirty archipelagos}
\label{sec:order30}

The order-thirty orbit provides a larger test, in which S-duality also permutes islands within a single archipelago. Its four symmetry descriptions are
\begin{equation}
 A=\mathrm{HET}(30D),\qquad B=\mathrm{HET}(-30D),\qquad
 C=\mathrm{HET}(-30E),\qquad D=\mathrm{II}(A_4+A_2),
\end{equation}
where the type II symmetry has Spin-lift order thirty. The frame-shape orbit predicts that $W_2$ exchanges
$A\leftrightarrow C$ and $B\leftrightarrow D$, while $W_3$ exchanges $A\leftrightarrow B$ and
$C\leftrightarrow D$. The operation $W_{15}$ preserves each of the four frame shapes. This last statement
leaves open a nontrivial action on the individual charge lattices.

\paragraph{Invariant lattices and shifts.}
The Leech fixed lattices have Gram matrices $\left(\begin{smallmatrix}8&2\\2&8\end{smallmatrix}\right)$, $\left(\begin{smallmatrix}4&2\\2&16\end{smallmatrix}\right)$, and $\left(\begin{smallmatrix}4&1\\1&4\end{smallmatrix}\right)$ for $A,B,C$, respectively. Their complements give two invariant-lattice classes for each of $A$ and $B$, and one for $C$. For $D$, the coinvariant lattice is $A_4\oplus A_2$, and the unique invariant lattice is likewise $I_1^{+,D}\simeq A_4\oplus A_2$. We continue to reverse the sign of the physical invariant lattices and kernels. Explicit bases and kernel representatives are given in Appendix~\ref{app:order30-data}.

\begin{center}
\begin{tabular}{ccccc}
\toprule
Description & \shortstack{Invariant-lattice\\classes} & $\det I^{+}$ & \shortstack{Shift classes\\per invariant lattice} & \shortstack{Kernels\\(Charge lattices)}\\
\midrule
$A$ & 2 & 60 & $218,\ 807$ & 40\\
$B$ & 2 & 60 & $539,\ 486$ & 40\\
$C$ & 1 & 15 & $274$ & 40\\
$D$ & 1 & 15 & $275$ & 40\\
\bottomrule
\end{tabular}
\end{center}
The kernel count identifies isometric kernels obtained from different shifts and, where applicable, from
different invariant lattices. Thus the four archipelagos each contain forty islands at the charge-lattice
level, despite their different numbers of shift presentations.

\paragraph{Kernels and genera.}
In each archipelago, the forty kernels form a complete genus. Consequently, all four collections can be specified compactly by their genus symbols:
\begin{equation}\label{eq:order30-genera}
 \begin{aligned}
 \mathcal G_A&=\mathrm{II}_{6,0}(2_{\mathrm{II}}^{+4}3^{+3}5^{+3}),&
 \mathcal G_B&=\mathrm{II}_{6,0}(2_{\mathrm{II}}^{+4}3^{-3}5^{-3}),\\
 \mathcal G_C&=\mathrm{II}_{6,0}(2_{\mathrm{II}}^{+2}3^{-3}5^{-3}),&
 \mathcal G_D&=\mathrm{II}_{6,0}(2_{\mathrm{II}}^{+2}3^{+3}5^{+3}).
 \end{aligned}
\end{equation}
The determinants are $54000$ for $\mathcal G_A,\mathcal G_B$ and $13500$ for $\mathcal G_C,\mathcal G_D$.
Completeness can again be checked by comparing the sum of reciprocal automorphism-group orders with the genus
mass. In particular, the four genera are distinct, so this orbit contains $160$ distinct kernel isometry
classes.

\paragraph{S-duality action.}
For each exact divisor $e\mid30$, the magnetic kernel transforms by the partial dual
\begin{equation}
 \mathcal D_e(K^{+})=\left(K^{+}+\frac{30}{e}(K^{+})^*\right)(e)
 =\left((K^{+})^*\cap\frac1e K^{+}\right)(e),
\end{equation}
with the electric lattice transforming to its dual. All kernels satisfy $30(K^{+})^*\subseteq K^{+}$. Let
$A_i$, $i=1,\ldots,40$, denote the representatives fixed in Appendix~\ref{app:order30-data}. We align the
labels in the other three genera by
\begin{equation}\label{eq:order30-alignment}
 B_i\simeq\mathcal D_3(A_i),\qquad
 C_i\simeq\mathcal D_2(A_i),\qquad
 D_i\simeq\mathcal D_6(A_i).
\end{equation}
Each of these maps is a bijection onto the forty kernels independently obtained by shift enumeration in the corresponding construction.

In this labelling, the action of $W_{15}$ is the same permutation in every archipelago:
\begin{equation}\label{eq:order30-permutation}
 \boxed{\sigma=(4\ 40)(5\ 38)(10\ 29)(11\ 17)(21\ 25).}
\end{equation}
All other indices are fixed. Thus $W_{15}$ fixes thirty islands and exchanges five pairs within each archipelago, although it fixes the frame shape throughout.

The complete action is given by the following table, with $j=\sigma(i)$:
\begin{center}
\begin{tabular}{c|ccccccc}
\toprule
 & $W_2$ & $W_3$ & $W_5$ & $W_6$ & $W_{10}$ & $W_{15}$ & $W_{30}$\\
\midrule
$A_i$ & $C_i$ & $B_i$ & $B_j$ & $D_i$ & $D_j$ & $A_j$ & $C_j$\\
$B_i$ & $D_i$ & $A_i$ & $A_j$ & $C_i$ & $C_j$ & $B_j$ & $D_j$\\
$C_i$ & $A_i$ & $D_i$ & $D_j$ & $B_i$ & $B_j$ & $C_j$ & $A_j$\\
$D_i$ & $B_i$ & $C_i$ & $C_j$ & $A_i$ & $A_j$ & $D_j$ & $B_j$\\
\bottomrule
\end{tabular}
\end{center}
Every entry is verified by integral isometry of the transformed kernel with its indicated target. In
particular, $W_3$ and $W_5$ have the same action on frame shapes but differ on ten islands in each
archipelago. Likewise, Fricke duality $W_{30}$ and $W_2$ connect the same pairs of archipelagos but need not
pair the same islands.

The full set therefore decomposes into thirty S-duality orbits of size four and five of size eight. A fixed
point of $\sigma$ gives the orbit $\{A_i,B_i,C_i,D_i\}$; a transposition $(i\ j)$ joins the two corresponding
quadruples into one orbit. There are $35$ S-duality classes in total at the charge-lattice level. This is the
nontrivial mixing that cannot be detected from frame shapes alone.

\section{Discussion of results}
\label{sec:discussion-results}

We have constructed a family of four-dimensional $\mathcal N=4$ string islands from cyclic orbifolds
of heterotic and type II strings, using symmetries of the Leech and $E_8$ lattices. For each symmetry,
we determined the admissible shifts and classified the resulting charge lattices. An ansatz for the
$R^2$ thresholds then predicts duality relations among these constructions, which we tested by explicit
comparison of their electromagnetic charge lattices under the assumptions of
Section~\ref{sec:charge-lattice-duality}. The agreement throughout the family provides evidence that
the perturbative duality pattern of CHL models extends to these islands.

These tests determine how the proposed Atkin--Lehner dualities act on the constructed archipelagos,
but leave open both the interpretation of additional strong-coupling limits and the extent of the family
of islands. Addressing these questions requires understanding exceptional cusps and possible discrete
theta angles, as well as extending the construction to non-cyclic orbifolds and symmetries beyond the
Conway construction.

\paragraph{Exceptional cusps.}
The threshold transformations are organized by the modular curve associated with $\Gamma_0(n)$.
Its cusps are equivalence classes of rational boundary points $\mathbb Q\cup\{\infty\}$, and a
weak-coupling description at a cusp would have $\operatorname{Im}S\to\infty$ in the corresponding
frame. At non-square-free orders, some cusps lie outside the Atkin--Lehner orbit of the known
weak-coupling limit. Our tests do not identify the physics at these exceptional cusps.

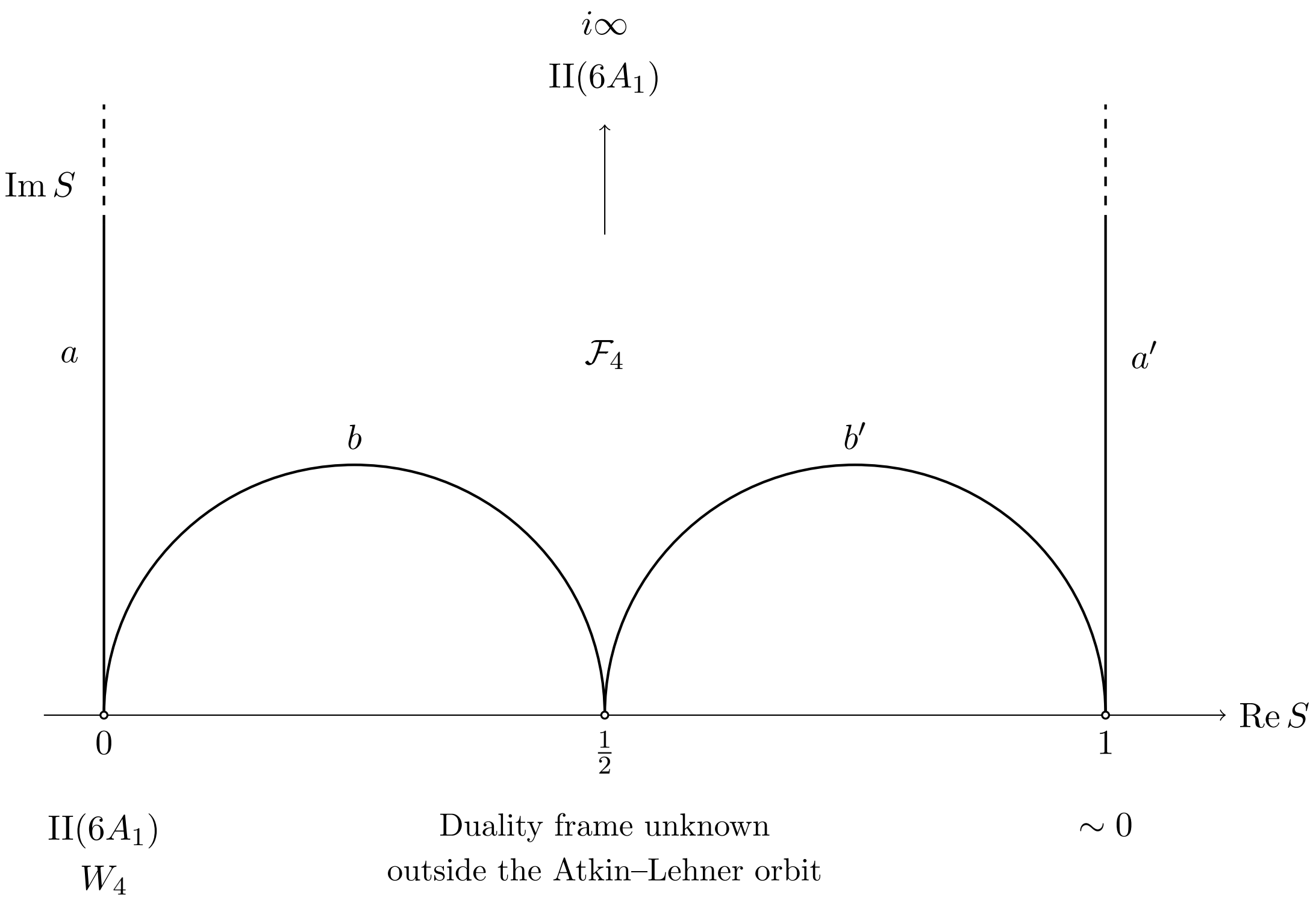
\begin{figure}[htbp]
\centering
\begin{tikzpicture}[x=11cm,y=11cm,font=\small]
\draw[->,thin] (-.06,0)--(1.12,0) node[right] {$\operatorname{Re}S$};
\draw[line width=.8pt] (0,0)--(0,.49) (1,0)--(1,.49);
\draw[dashed,line width=.8pt] (0,.49)--(0,.61) (1,.49)--(1,.61);
\node[anchor=east] at (-.015,.53) {$\operatorname{Im}S$};
\node[anchor=east] at (-.012,.36) {$a$};
\node[anchor=west] at (1.012,.36) {$a^{\prime}$};
\node at (.5,.36) {$\mathcal F_{4}$};
\draw[->,thin] (.5,.48)--(.5,.59);
\node[align=center,anchor=south] at (.5,.605) {$i\infty$\\[3pt]$\mathrm{II}(6A_1)$};
\draw[line width=.8pt] (0.00000000,0) arc[start angle=180,end angle=0,radius=0.25000000];
\node[inner sep=1pt,anchor=south] at (0.25000000,0.26200000) {$b$};
\draw[line width=.8pt] (0.50000000,0) arc[start angle=180,end angle=0,radius=0.25000000];
\node[inner sep=1pt,anchor=south] at (0.75000000,0.26200000) {$b^{\prime}$};
\draw[fill=white,line width=.6pt] (0.00000000,0) circle[radius=.0035];
\node[anchor=north,inner sep=3pt] at (0.00000000,-.006) {$0$};
\draw[fill=white,line width=.6pt] (0.50000000,0) circle[radius=.0035];
\node[anchor=north,inner sep=3pt] at (0.50000000,-.006) {$\frac{1}{2}$};
\draw[fill=white,line width=.6pt] (1.00000000,0) circle[radius=.0035];
\node[anchor=north,inner sep=3pt] at (1.00000000,-.006) {$1$};
\node[align=center,anchor=north] at (0,-.085) {$\mathrm{II}(6A_1)$\\[2pt]$W_4$};
\node[align=center,anchor=north,font=\footnotesize] at (.5,-.085) {Duality frame unknown\\[2pt]outside the Atkin--Lehner orbit};
\node[anchor=north] at (1,-.085) {$\sim0$};
\end{tikzpicture}
\caption{A fundamental domain $\mathcal F_4$ for $\Gamma_0(4)$ for the type II $\mathbb Z_4$ island. The region lies between the vertical sides and above the two semicircles. Matching letters denote identified sides, and $0\sim1$. The three cusp classes are therefore $\infty$, $0$, and $1/2$. Fricke duality $W_4$ exchanges the two known type II limits at infinity and zero. The cusp $1/2$ lies outside this Atkin--Lehner orbit, and its string interpretation remains open.}
\label{fig:order4-axiodilaton}
\end{figure}

The simplest example is the type IIA $\mathbb Z_4$ island, illustrated in
Figure~\ref{fig:order4-axiodilaton}. The cusps of $\Gamma_0(4)$ have representatives
$\infty$, $0$, and $1/2$. Fricke duality $W_4$ exchanges the first two, while $1/2$ lies outside
their orbit; there is no $W_2$, since two is not an exact divisor of four. It remains to determine
whether this limit admits a perturbative string description. In particular, the exceptional cusps
of the order-four and order-eight islands are possible locations for heterotic frames.

This issue already arises for the standard $\mathbb Z_4$ CHL model. Its balanced frame shape
$1^4 2^2 4^4$ admits Fricke self-duality, but the cusp $1/2$ remains outside the orbit of $\infty$.
This cusp appears explicitly in the modular-form analysis of Appendix~A.6
of~\cite{Govindarajan:2009qt}. Understanding its physical interpretation could help identify the
corresponding limits of the islands.

\paragraph{Models with nontrivial level-matching anomaly.}
The models with $\lambda>1$ remain to be studied. Their shifts must compensate the level mismatch of the
symmetry, and the CHL analysis indicates that their electromagnetic charge lattices require additional
data~\cite{Persson:2015jka}. Extending our S-duality tests to these models therefore requires determining
their full charge lattices, including possible nonperturbative electric charges, and the appropriate
electric--magnetic decomposition. We leave this extension for future work.

\paragraph{Discrete theta angles.}
Many of the type II models appearing in CHL constructions admit discrete theta
angles~\cite{Baykara:2025lhl,Montero:2022vva}, and we expect analogous choices for some of the
islands studied here. It remains to determine which constructions admit them and how S-duality
acts on these choices. Their inclusion may refine the identification of dual theories, and would
require extending the present charge-lattice and threshold analysis to account for the discrete data.

\paragraph{Non-cyclic orbifolds.}
Our constructions use a single symmetry generator together with a shift. Non-cyclic orbifolds
provide a further direction even when all generators are drawn from the lattice symmetry groups
used here. Constructing such models requires compatible group actions and shifts, with consistency
conditions imposed on the full orbifold group. The resulting islands could furnish additional duality
frames, potentially including descriptions at the exceptional cusps. This possibility requires an
explicit construction and a comparison with the predicted charges and thresholds.

\paragraph{Symmetries beyond Conway.}
On the heterotic side, we have used coinvariant lattices inherited from the Leech lattice, rather
than classified all admissible symmetries of the Narain lattice. More general integral realizations
could therefore produce islands outside the Conway construction. The lattice formulation gives a
concrete starting point for this extension: one must find rootless rank-twenty-two coinvariant lattices
with suitable automorphisms and primitive embeddings into $\Gamma_{22,6}$, as described in
Section~\ref{sec:heterotic-construction}. Determining which of these possibilities yield consistent
islands, and whether they enlarge the duality orbits found here, is an open classification problem.

\section*{Acknowledgements}

We thank Anamaria Font, Miguel Montero and Ida Zadeh for many useful discussions and correspondence. We thank
the Simons Center for Geometry and Physics for hospitality during the 2026 Simons Physics Summer Workshop. HCT would like to thank the Harvard Swampland Initiative for its hospitality during part of this work. The work of HP is supported by a grant from the Simons Foundation (602883,CV), the DellaPietra Foundation, the
NSF grant PHY2013858 and by a UKRI Future Leaders Fellowship (UKRI2349).

\noindent We used OpenAI Codex to assist with manuscript editing. The authors take full responsibility for the final manuscript.

\appendix

\section{Lattice conventions, embeddings and genera}
\label{app:lattice-definitions}

\subsection{Basic definitions}
An integral lattice $L$ is a free abelian group of finite rank equipped with a nondegenerate symmetric
bilinear pairing $(\cdot,\cdot):L\times L\to\mathbb Z$. Its Gram matrix in a basis $e_i$ is
$G_{ij}=(e_i,e_j)$. Its signature $(r_+,r_-)$ counts positive and negative eigenvalues of this matrix. We call
a positive-definite lattice Euclidean. The lattice is even if $(x,x)\in2\mathbb Z$ for every $x\in L$. Its
dual is
\begin{equation}
 L^*=\{x\in L\otimes\mathbb Q:(x,L)\subseteq\mathbb Z\}.
\end{equation}
An integral lattice is unimodular, or self-dual, when $L=L^*$, equivalently $|\det G|=1$. We write $\det
L=\det G$, and $L(c)$ for the same abelian group with its pairing multiplied by $c$. In particular, $L(-1)$
reverses the signature. The group $O(L)$ consists of integral automorphisms preserving the pairing. Lattice
isometry means equivalence under an integral change of basis of determinant $\pm1$.

The discriminant group and, for even $L$, its discriminant quadratic form are
\begin{equation}
 A_L=L^*/L,\qquad |A_L|=|\det L|,\qquad
 q_L:A_L\longrightarrow\mathbb Q/2\mathbb Z,\quad
 q_L([x])=(x,x)\bmod2\mathbb Z.
\end{equation}
An isomorphism of discriminant forms preserves both the finite group and this quadratic function. We denote by
$\ell(A_L)$ the minimum number of generators of $A_L$, which is at most $\operatorname{rank}L$.

In our positive-definite normalization, a root is a vector of squared length two. A rootless lattice contains no such vectors. The symbols $A_r,D_r,E_r$ denote the corresponding root lattices in this normalization. The $E_8$ lattice is the unique even unimodular Euclidean lattice of rank eight. The Leech lattice is the unique rootless even unimodular Euclidean lattice of rank twenty-four. The hyperbolic plane $U=\Gamma_{1,1}$ has Gram matrix $\left(\begin{smallmatrix}0&1\\1&0\end{smallmatrix}\right)$.

\subsection{Primitive embeddings and gluing}
\label{app:lattice-gluing}
An embedding $L\hookrightarrow M$ is primitive when $M/L$ has no torsion, equivalently
\begin{equation}
 (L\otimes\mathbb Q)\cap M=L.
\end{equation}
For any sublattice $R\subset M$, its saturation is $(R\otimes\mathbb Q)\cap M$. An overlattice of $L$ is a
lattice containing $L$ with finite index and the same pairing. If $L\subset M$ has index $d$, then $|\det
L|=d^2|\det M|$.

Suppose that even lattices $N$ and $I$ have opposite discriminant forms. Choose an anti-isometry $\phi:A_N\to
A_I$, meaning $q_I(\phi(a))=-q_N(a)$. Their gluing is
\begin{equation}\label{eq:gluing-definition}
 \Gamma_\phi=\{(x,y)\in N^*\oplus I^*:[y]=\phi([x])\}.
\end{equation}
This is an even unimodular overlattice of $N\oplus I$, with primitive orthogonal sublattices $N$ and $I$.
Conversely, primitive orthogonal complements in an even unimodular lattice have opposite discriminant
forms~\cite{NikulinIntegralForms}. Thus gluing identifies fractional vectors in the two dual lattices whose
combined quadratic norm is even.

For embeddings of abstract lattices, gluing maps related by the induced actions of $O(N)$ and $O(I)$ give
equivalent embeddings. If a specified symmetry $\Theta$ is retained, the equivalence must also intertwine its
action, restricting the allowed automorphisms on a fixed $N$. When $\Theta$ acts trivially on
$A_N$, the action $\Theta\oplus1$ preserves~\eqref{eq:gluing-definition}. This is the extension criterion used
in the main text.

\subsection{Rank-two embeddings into \texorpdfstring{$E_8$}{E8}}
\label{app:rank-two-E8}
Every even positive-definite lattice $L$ of rank two admits a primitive embedding into $E_8$. The
qualifications ``even'' and ``positive definite'' are essential. Here is a short proof using Nikulin's
existence theorem for even lattices with prescribed signature and discriminant form, Theorem~1.10.1
of~\cite{NikulinIntegralForms}.

The group $A_L$ has length at most two. The signature of its finite quadratic form is $2\bmod8$, so the
opposite form $-q_L$ has signature $6\bmod8$. Here the signature of a discriminant form is the signature
difference $r_+-r_-$, modulo eight, of any even lattice realizing it. Nikulin's theorem therefore supplies an
even lattice $I^{+}$ of signature $(6,0)$ and discriminant form $-q_L$. The signature congruence holds, and
its rank six is strictly greater than the length of every primary part of $A_L$, so the additional local
determinant conditions at equality do not arise. Gluing $L$ and $I^{+}$ along their opposite discriminant
forms gives an even unimodular positive-definite lattice of rank eight. This lattice is $E_8$, and the gluing
construction makes $L$ primitive with orthogonal complement $I^{+}$. This proves existence; different
complement classes can give inequivalent embeddings.

\subsection{Genera, genus symbols and mass}
\label{app:lattice-genera}
Two integral lattices are in the same genus if they are isometric over $\mathbb R$ and over the $p$-adic
integers $\mathbb Z_p$ for every prime $p$. The latter comparison retains integral information at each prime,
including congruences modulo all powers of $p$. Lattices in one genus can nevertheless be nonisometric over
$\mathbb Z$. For even lattices, the genus is determined by the signature and discriminant quadratic
form~\cite{NikulinIntegralForms}. A complete genus in our tables means the full set of integral isometry
classes in that genus. Its class number is the number of those classes.

A genus symbol records the signature and local Jordan data. Locally one decomposes
\begin{equation}
 L\otimes\mathbb Z_p\simeq\bigoplus_{a\geq0}L_{p,a}(p^a),
\end{equation}
where each $L_{p,a}$ is unimodular over $\mathbb Z_p$. The prefix $\mathrm{II}_{r_+,r_-}$ denotes an even
lattice of that signature. For odd $p$, a symbol $(p^a)^{\epsilon r}$ records a component of rank $r$, with
$\epsilon=+$ or $-$ according to whether the determinant of its unimodular part is a quadratic residue or
nonresidue modulo $p$. At $p=2$, the sign is positive for determinant $\pm1\bmod8$ and negative for
$\pm3\bmod8$. The subscript $\mathrm{II}$ specifies an even unimodular part; for an odd part, a numerical
subscript gives its oddity, the sum modulo eight of the odd diagonal units in a diagonalization. Unimodular
components can be left implicit, as in the tables. A genus can have several equivalent symbols arising from
different local decompositions.

For a positive-definite genus $\mathcal G$, the genus mass is
\begin{equation}\label{eq:genus-mass-definition}
 \operatorname{mass}(\mathcal G)
 =\sum_{[L]\in\mathcal G}\frac{1}{|O(L)|},
\end{equation}
where the sum runs once over each integral isometry class. The Siegel--Minkowski mass formula computes this
quantity from local data. Consequently, a list of pairwise nonisometric lattices known to belong to $\mathcal
G$ is complete if the sum of their reciprocal automorphism-group orders equals the genus mass. This is the
completeness check referred to in the examples. The mass is a weighted count, whereas the class number counts
every class once.

Finally, a positive-definite lattice is $n$-modular when $L\simeq L^*(n)$. For rank $r$, a necessary condition
is $\det L=n^{r/2}$. In the rank-six examples the required determinant is therefore $n^3$.

\section{Enumerating shifts and their charge lattices}
\label{app:shift-enumeration}

We describe the finite calculation used in the examples, working with the positive-definite lattice
$I^{+}=I(-1)$. Choose a basis with Gram matrix $G$ and write a shift as $v=w/n$, where $w$ is an integral row
vector. The integrality conditions and the kernel of the shift character are invariant under this sign
reversal.

\paragraph{From shifts to a finite list.}
Since adding a vector of $I^{+}$ to $v$ does not change the shift class, we may first take every entry of $w$
between zero and $n-1$. This gives a finite list of $n^6$ candidates. Level matching requires
\begin{equation}
    \frac{wGw^{\mathsf T}}{2n}\in\mathbb Z.
\end{equation}
For each surviving candidate, compute the six residues
\begin{equation}\label{eq:shift-character-coordinates}
    c=wG\pmod n.
\end{equation}
A lattice vector with integral coordinates $x$ acquires
the phase $\exp(2\pi i\,xc^{\mathsf T}/n)$. We retain shifts for which this character has exact order $n$,
equivalently
\begin{equation}
    \gcd(n,c_1,\ldots,c_6)=1.
\end{equation}
This is the sufficient condition used in the main text to exclude massless states in every nontrivial twisted sector.

\paragraph{Removing equivalent descriptions.}
Two numerator vectors with the same residues $c$ differ by a translation of $v$ by $(I^{+})^*$, so they give
the same shift character. Multiplying $c$ by an integer coprime to $n$ changes the generator of the cyclic
orbifold and is also identified. Finally, we identify shifts related by lattice automorphisms. In row-vector conventions, an
integral matrix $A$ satisfying $AGA^{\mathsf T}=G$ sends $w$ to $wA$. We keep one representative of each
resulting class among the allowed candidates.

\paragraph{The order-twenty-three example.}
For $n=23$, arithmetic modulo $23$ is arithmetic in a field. The matrix $G_{I^{+}}$
in~\eqref{eq:order23-invariant} has rank five modulo $23$. There is therefore one independent direction $r$
for which $rG_{I^{+}}=0\pmod{23}$. Adding any multiple of $r$ to $w$ leaves the shift phases unchanged. This
set of invisible directions is called the \emph{radical}. After removing it, only five independent residues
remain.

In these five coordinates, level matching becomes the condition that the quadratic norm vanish modulo $23$. A
nonzero vector satisfying this modular norm condition is called \emph{isotropic} modulo $23$. All its nonzero
multiples give equivalent generators. The set of these multiples is what is meant by an \emph{isotropic line}.
Thus counting isotropic lines simply means listing nonzero solutions of the modular norm condition and
identifying scalar multiples. There are $23^3+23^2+23+1=12720$ such classes before lattice automorphisms are
imposed. The $480$ automorphisms of $I^{+}$ reduce them to $84$ shift classes. Their multiplicities
among the $29$ kernel classes are recorded in Table~\ref{tab:order23-kernels}.

For a composite order, one can instead use the residue calculation~\eqref{eq:shift-character-coordinates}
directly. For example, the changes of generator modulo nine are multiplication by $1,2,4,5,7,8$.

\paragraph{Constructing and comparing the kernels.}
For each shift, solve the single congruence
\begin{equation}
    K^{+}_w=\{x\in\mathbb Z^6:xc^{\mathsf T}=0\pmod n\}.
\end{equation}
If $c_a$ is invertible modulo $n$, an explicit basis consists of $ne_a$ and the five vectors $e_j-b_je_a$,
with $b_j\equiv c_jc_a^{-1}\pmod n$ and $j\neq a$. Otherwise one can solve the same congruence using integer
row reduction. If the basis vectors are the rows of $B$, the kernel Gram matrix is $BGB^{\mathsf T}$, with
determinant $n^2\det G$. We compare these matrices by integral isometry, which can identify kernels arising
from inequivalent shifts. The final check of Fricke duality is an integral isometry between a kernel Gram
matrix in one theory and $n$ times the inverse kernel Gram matrix in its proposed dual.

\paragraph{The order-nine calculation.}
For each Gram matrix in~\eqref{eq:order9-invariants}, one can run through all $9^6$ numerator vectors
explicitly. For the heterotic lattice, $69984$ vectors satisfy level matching and the exact-order condition.
Identifying equal characters and multiplication by units leaves $432$ classes. Lattice automorphisms divide
these into three orbits, of sizes $144$, $96$, and $192$, represented in that order by the three heterotic
numerators in~\eqref{eq:order9-shifts}. All three kernels are isometric. For the type II lattice, the
corresponding counts are $58320$ numerator vectors and $3240$ character classes modulo units, which form a
single orbit under $O(E_6)$.

For example, $w_{\mathrm{het},1}$ gives $c=(8,7,7,4,4,5)\pmod9$. Since $8^{-1}=8\pmod9$, a kernel basis is
$9e_1$, $e_2-2e_1$, $e_3-2e_1$, $e_4-5e_1$, $e_5-5e_1$, and $e_6-4e_1$. An integral change of basis gives the
smaller matrix entries displayed in $G_{K^{+,\mathrm{het}}}$ in~\eqref{eq:order9-kernels} without changing the lattice. This illustrates that the kernel is obtained simply by solving the condition for the shift phase
to be trivial.

Finally, for the kernel matrices displayed in~\eqref{eq:order9-kernels}, the unimodular matrix
\begin{equation}
 U=\begin{pmatrix}
 0&0&0&0&0&-1\\
 0&0&0&0&-1&0\\
 0&0&0&-1&0&0\\
 0&0&-1&0&1&1\\
 0&1&0&0&0&0\\
 -1&0&0&1&1&1
 \end{pmatrix}
 \quad\text{satisfies}\quad
 U\bigl(9G_{K^{+,\mathrm{het}}}^{-1}\bigr)U^{\mathsf T}
 =G_{K^{+,\mathrm{II}}}.
\end{equation}
This supplies a direct certificate for the Fricke pairing, including the integral structure of the charge lattices.

\paragraph{The order-six calculation.}
The same procedure, now testing $6^6$ numerator vectors, gives $8640$ allowed vectors for $I_D^{+}$
in~\eqref{eq:order6-invariant}. Identifying equal characters and multiplication by the units $1,5$ leaves
$360$ classes. The automorphism group has two orbits, of sizes $216$ and $144$, represented by $u_1$ and
$u_2$; their kernels are isometric. For $I_E^{+}=E_6$, there are $6480$ allowed vectors and $1080$ character
classes modulo units, forming one automorphism orbit. These two calculations apply to all four symmetry
descriptions in Section~\ref{sec:order6}.

To check $W_e$ for $e=2,3,6$, one can also construct~\eqref{eq:order6-partial-dual} directly from the kernel
Gram matrix $G$. In coordinates relative to a dual basis, $K^{+}$ is generated by the rows of $G$, while
$(6/e)(K^{+})^*$ is generated by the rows of $(6/e)\mathbf1_6$. Integer row reduction of these twelve
generators gives a basis matrix $B_e$ for their sum. Its rescaled Gram matrix is $eB_eG^{-1}B_e^{\mathsf T}$.
Comparing this matrix with the two matrices in~\eqref{eq:order6-kernels} by integral isometry verifies all six
relations in~\eqref{eq:order6-dualities}.

\section{Explicit lattice and shift data}
\label{app:explicit-data}

We collect the longer data tables for the examples of Section~\ref{sec:explicit-results}. Throughout,
$I^{+}=I(-1)$ and $K^{+}=(\ker_Iv)(-1)$ are positive definite. A shift is written $v=w/n$, with the integral
row vector $w$ expressed in the specified basis of $I^{+}$, and its kernel is
\begin{equation}
 K^{+}_w=\{x\in\mathbb Z^6:xG_{I^{+}}w^{\mathsf T}=0\pmod n\}.
\end{equation}
Kernel indices label isometry classes. A shift multiplicity counts the inequivalent shift classes defined
in Appendix~\ref{app:shift-enumeration} that yield a given kernel, without assuming that they define
inequivalent CFTs. In both examples we specify the invariant basis and a defining shift for each kernel,
together with the shift multiplicities. The shorter order-four, order-eight, order-nine and order-six data are displayed
directly in their respective worked examples. The complete list of genera is in
Section~\ref{sec:general-results}.

\subsection{Order twenty-three: kernel representatives and shift multiplicities}
\label{app:order23-shifts}
All vectors below are numerator coordinates $w$ for the shift $v=w/23$ in the basis
of~\eqref{eq:order23-invariant}. Table~\ref{tab:order23-kernels} specifies one representative for each kernel
and its multiplicity, minimum, and automorphism order.

\begingroup
\small
\renewcommand{\baselinestretch}{1}\selectfont
\setlength{\tabcolsep}{7pt}
\renewcommand{\arraystretch}{1.12}
\begin{longtable}{rcrrr}
\caption[Order-twenty-three kernels]{The twenty-nine order-twenty-three kernels. The vector $w_i$ specifies $v_i=w_i/23$; $d_i$ counts all shift classes giving $K^{+}_i$. The minimum is the minimum squared norm, and $|O(K_i^{+})|$ is the lattice automorphism-group order. Every row is fixed by $W_{23}$.}\label{tab:order23-kernels}\\
\toprule
$i$ & $w_i$ & $d_i$ & $\min K^{+}_i$ & $|O(K^{+}_i)|$\\
\midrule
\endfirsthead
\multicolumn{5}{c}{Table \thetable\ (continued)}\\
\toprule
$i$ & $w_i$ & $d_i$ & $\min K^{+}_i$ & $|O(K^{+}_i)|$\\
\midrule
\endhead
\bottomrule
\endfoot
1 & $(12,2,1,0,0,0)$ & 6 & 2 & 16 \\
2 & $(8,1,1,1,0,0)$ & 2 & 2 & 288 \\
3 & $(6,2,1,1,0,0)$ & 1 & 2 & 384 \\
4 & $(4,7,1,1,0,0)$ & 3 & 2 & 64 \\
5 & $(1,9,1,1,0,0)$ & 3 & 2 & 64 \\
6 & $(3,4,2,1,0,0)$ & 12 & 2 & 8 \\
7 & $(2,6,4,1,0,0)$ & 6 & 2 & 16 \\
8 & $(13,2,1,0,1,0)$ & 3 & 2 & 32 \\
9 & $(8,3,1,0,1,0)$ & 1 & 4 & 48 \\
10 & $(2,7,1,0,1,0)$ & 3 & 4 & 16 \\
11 & $(2,8,2,0,1,0)$ & 5 & 4 & 12 \\
12 & $(6,9,4,0,1,0)$ & 1 & 6 & 48 \\
13 & $(4,11,4,0,1,0)$ & 2 & 4 & 32 \\
14 & $(12,1,1,1,1,0)$ & 2 & 2 & 144 \\
15 & $(14,3,2,1,1,0)$ & 4 & 4 & 12 \\
16 & $(1,7,2,1,1,0)$ & 3 & 4 & 16 \\
17 & $(2,11,2,1,1,0)$ & 2 & 2 & 48 \\
18 & $(5,12,2,1,1,0)$ & 6 & 4 & 8 \\
19 & $(3,3,3,1,1,0)$ & 6 & 2 & 16 \\
20 & $(5,6,5,1,1,0)$ & 2 & 4 & 24 \\
21 & $(4,8,7,1,1,0)$ & 1 & 6 & 48 \\
22 & $(3,16,15,1,1,0)$ & 1 & 4 & 48 \\
23 & $(4,5,3,2,1,0)$ & 1 & 4 & 48 \\
24 & $(7,11,5,2,1,0)$ & 2 & 6 & 24 \\
25 & $(6,10,6,2,1,0)$ & 1 & 6 & 48 \\
26 & $(4,18,11,2,1,0)$ & 1 & 2 & 96 \\
27 & $(9,9,7,3,1,0)$ & 1 & 4 & 48 \\
28 & $(5,18,12,3,1,0)$ & 2 & 2 & 144 \\
29 & $(16,4,4,4,1,0)$ & 1 & 2 & 1152 \\
\end{longtable}
\endgroup

\subsection{Order thirty: invariant lattices, kernel representatives, and shift multiplicities}
\label{app:order30-data}

This subsection fixes the bases and kernel numbering used in Section~\ref{sec:order30}, with the conventions
of Appendix~\ref{app:explicit-data}. Write $I_a^{+,X}$ for the $a$-th positive-definite invariant lattice of
description $X=A,B,C,D$. The six Gram matrices are
\begin{equation}
G_{A,1}=\begin{pmatrix}
2&-1&-1&0&0&1\\
-1&2&0&0&0&0\\
-1&0&2&0&0&0\\
0&0&0&2&-1&0\\
0&0&0&-1&2&0\\
1&0&0&0&0&6
\end{pmatrix},\qquad G_{A,2}=\begin{pmatrix}
2&0&0&0&1&0\\
0&2&-1&1&1&0\\
0&-1&2&0&-1&0\\
0&1&0&4&2&0\\
1&1&-1&2&4&-1\\
0&0&0&0&-1&2
\end{pmatrix}.
\end{equation}
\begin{equation}
G_{B,1}=\begin{pmatrix}
2&-1&1&0&1&1\\
-1&2&-1&0&-1&-1\\
1&-1&2&0&0&0\\
0&0&0&2&0&1\\
1&-1&0&0&4&1\\
1&-1&0&1&1&4
\end{pmatrix},\qquad G_{B,2}=\begin{pmatrix}
2&-1&0&1&0&0\\
-1&2&0&-1&0&0\\
0&0&4&-2&0&0\\
1&-1&-2&4&-1&-1\\
0&0&0&-1&2&1\\
0&0&0&-1&1&2
\end{pmatrix}.
\end{equation}
\begin{equation}
G_{C,1}=\begin{pmatrix}
2&1&1&1&1&-1\\
1&2&1&1&1&-1\\
1&1&2&1&1&0\\
1&1&1&2&1&0\\
1&1&1&1&2&-1\\
-1&-1&0&0&-1&4
\end{pmatrix},\qquad G_{D,1}=\begin{pmatrix}
2&-1&0&0&0&0\\
-1&2&-1&0&0&0\\
0&-1&2&-1&0&0\\
0&0&-1&2&0&0\\
0&0&0&0&2&-1\\
0&0&0&0&-1&2
\end{pmatrix}.
\end{equation}

In the displayed order, their automorphism-group orders are $1152,192,384,288,2880,2880$. The pairs for $A$
and $B$ each exhaust their respective invariant-lattice genera, while those for $C$ and $D$ are single-class
genera.

\paragraph{Enumerating the shifts.}
To avoid testing all $30^6$ numerator vectors directly, one can first solve the level-matching condition
$wGw^{\mathsf T}/2=0$ separately modulo two, three, and five. At each prime, retain a nonzero character $wG$
and identify its nonzero scalar multiples. The Chinese remainder theorem then combines any choice of these
three residue classes into a unique character modulo thirty, up to a change of generator. There are
$9\cdot40\cdot156=56160$ such classes for each invariant lattice of $A$ and $B$, and
$35\cdot40\cdot156=218400$ for each of $C$ and $D$. Taking lattice-automorphism orbits gives the shift counts
in the main text. The kernels are constructed and compared as in Appendix~\ref{app:shift-enumeration}.

\paragraph{Fixing the kernel labels.}
The following table defines all forty $A_i$ by a numerator $w$ and a choice $a$ of invariant lattice:
\begin{equation}
 A_i=\{x\in\mathbb Z^6:xG_{A,a}w^{\mathsf T}=0\pmod{30}\},
 \qquad (x,y)=xG_{A,a}y^{\mathsf T}.
\end{equation}
The other $120$ kernels are specified by the partial-dual relations~\eqref{eq:order30-alignment}. The columns
$\nu_X$ count the inequivalent shifts yielding $X_i$, summed over its invariant lattices. The
automorphism-group order is the same for all four kernels with aligned index $i$. Thus the table also records
how the different numbers of shift presentations collapse to forty charge-lattice classes in each
construction.

\begingroup
\small
\setlength{\tabcolsep}{4pt}
\begin{longtable}{rrlrrrrr}
\caption{Order-thirty charge-kernel representatives and shift multiplicities. The index $i$ labels the kernel, $a$ selects the invariant Gram matrix $G_{A,a}$, and $v=w/30$. Each $\nu_X$ counts the shifts giving the aligned kernel in description $X$, summed over invariant lattices. The final column gives the common lattice automorphism-group order.}\label{tab:order30-kernels}\\
\toprule
$i$&$a$&$w$&$\nu_A$&$\nu_B$&$\nu_C$&$\nu_D$&$|O(A_i)|$\\
\midrule
\endfirsthead
\multicolumn{8}{c}{Table \thetable\ (continued)}\\
\toprule
$i$&$a$&$w$&$\nu_A$&$\nu_B$&$\nu_C$&$\nu_D$&$|O(A_i)|$\\
\midrule
\endhead
\bottomrule
\endfoot
1 & 1 & $(17,10,0,21,0,0)$ & 8 & 12 & 2 & 3 & 256 \\
2 & 1 & $(5,22,6,15,0,0)$ & 12 & 12 & 2 & 2 & 256 \\
3 & 1 & $(3,16,16,25,0,0)$ & 20 & 24 & 4 & 5 & 64 \\
4 & 1 & $(23,16,6,15,0,0)$ & 26 & 14 & 5 & 2 & 64 \\
5 & 1 & $(5,10,12,21,0,0)$ & 60 & 64 & 16 & 10 & 16 \\
6 & 1 & $(5,22,0,21,0,0)$ & 18 & 22 & 4 & 5 & 64 \\
7 & 1 & $(15,22,10,1,0,0)$ & 4 & 5 & 2 & 2 & 576 \\
8 & 1 & $(23,16,0,27,6,0)$ & 9 & 8 & 2 & 3 & 128 \\
9 & 1 & $(11,22,0,3,6,0)$ & 10 & 8 & 2 & 2 & 256 \\
10 & 1 & $(29,16,12,15,6,0)$ & 60 & 27 & 8 & 6 & 16 \\
11 & 1 & $(11,10,12,3,6,0)$ & 8 & 4 & 2 & 2 & 384 \\
12 & 1 & $(15,22,10,25,6,0)$ & 20 & 22 & 8 & 10 & 32 \\
13 & 1 & $(9,22,16,25,6,0)$ & 9 & 12 & 4 & 6 & 64 \\
14 & 1 & $(5,22,0,15,6,0)$ & 15 & 18 & 4 & 6 & 32 \\
15 & 1 & $(5,10,12,15,6,0)$ & 60 & 72 & 16 & 24 & 8 \\
16 & 1 & $(3,10,10,25,6,0)$ & 16 & 18 & 8 & 8 & 64 \\
17 & 1 & $(15,4,4,7,6,0)$ & 4 & 9 & 4 & 4 & 384 \\
18 & 1 & $(5,16,6,27,6,0)$ & 20 & 28 & 8 & 10 & 32 \\
19 & 1 & $(9,12,6,21,0,10)$ & 42 & 54 & 6 & 6 & 16 \\
20 & 1 & $(3,6,6,15,0,10)$ & 30 & 24 & 5 & 4 & 64 \\
21 & 1 & $(5,2,0,21,0,10)$ & 15 & 18 & 7 & 4 & 96 \\
22 & 1 & $(15,12,0,21,0,10)$ & 9 & 10 & 2 & 2 & 96 \\
23 & 1 & $(27,6,0,3,6,10)$ & 27 & 18 & 3 & 3 & 32 \\
24 & 1 & $(5,8,12,3,6,10)$ & 3 & 2 & 1 & 1 & 2304 \\
25 & 1 & $(15,12,12,3,6,10)$ & 14 & 8 & 3 & 3 & 96 \\
26 & 1 & $(9,12,12,3,6,10)$ & 12 & 8 & 3 & 2 & 128 \\
27 & 1 & $(13,4,24,7,6,10)$ & 12 & 9 & 8 & 4 & 72 \\
28 & 1 & $(3,6,0,27,6,10)$ & 108 & 108 & 24 & 24 & 4 \\
29 & 1 & $(5,2,0,15,6,10)$ & 36 & 56 & 12 & 18 & 16 \\
30 & 1 & $(15,6,6,27,6,10)$ & 38 & 34 & 14 & 12 & 16 \\
31 & 1 & $(9,12,6,15,6,10)$ & 54 & 54 & 12 & 12 & 8 \\
32 & 1 & $(27,0,0,15,6,10)$ & 72 & 63 & 24 & 18 & 8 \\
33 & 1 & $(9,12,0,27,6,10)$ & 27 & 27 & 6 & 6 & 16 \\
34 & 1 & $(7,10,0,25,6,10)$ & 36 & 33 & 12 & 10 & 16 \\
35 & 1 & $(15,12,0,15,6,10)$ & 32 & 28 & 10 & 8 & 16 \\
36 & 1 & $(1,28,12,25,6,10)$ & 19 & 19 & 4 & 4 & 24 \\
37 & 1 & $(27,6,6,27,6,10)$ & 12 & 10 & 6 & 4 & 64 \\
38 & 1 & $(13,10,0,25,6,10)$ & 29 & 38 & 6 & 12 & 16 \\
39 & 2 & $(10,0,12,10,15,6)$ & 9 & 9 & 2 & 2 & 48 \\
40 & 2 & $(22,0,6,10,15,0)$ & 10 & 16 & 3 & 6 & 64 \\
\midrule
\multicolumn{3}{r}{Total shifts}&1025&1025&274&275&\\
\end{longtable}
\endgroup

For every kernel and each $e=2,3,5,6,10,15,30$, we construct a basis of $K^{+}+(30/e)(K^{+})^*$ by integer row
reduction, exactly as in the order-six example, and test the resulting Gram matrix against the target
construction. All $160\cdot7=1120$ comparisons admit integral unimodular isometries.

\section{Comparison of the modular kernels with the lattice catalogue}
\label{app:modular-catalogue}

We checked every kernel in our classification for an integral isometry
$K^{+}\simeq (K^{+})^*(m)$, where rank six requires $\det K^{+}=m^3$.
After identifying repeated abstract lattices across descriptions, there are
$121$ distinct modular classes. Here $m$ is the modularity parameter,
determined by the lattice isometry above. In the cases below
it equals the orbifold order. The count concerns invariance of the lattice under Fricke duality. Self-duality of a description also requires invariance of its threshold frame shape. In particular, the
three $12$-modular classes occur in the $-12D$ orbit, whose two descriptions are exchanged by Fricke duality,
as well as in the $-12K$ constructions.

\begin{center}
\begin{tabular}{crrl}
\toprule
\shortstack{Modularity\\$m$}&\shortstack{Lattice\\classes}&\shortstack{Catalogue\\matches}&Catalogue identification\\
\midrule
4&1&1&Single-class list, rank six, entry 155\\
8&4&1&\texttt{6QF.6.d}\quad($K^{+}_{8,3}$)\\
12&3&0&---\\
15&18&0&---\\
16&2&0&---\\
20&13&0&---\\
23&29&0&---\\
24&38&0&---\\
28&13&0&---\\
\midrule
Total&121&2&\\
\bottomrule
\end{tabular}
\end{center}

The comparison uses the six-dimensional listings and modular-lattice tables
of the Nebe--Sloane catalogue~\cite{NebeSloaneCatalogue}, its downloadable
standard records, and the linked list of primitive single-class genera.
The downloaded files were checked on 22 November 2026. We compared Gram
matrices by integral isometry, allowing a common rational rescaling of a
catalogue matrix when it gives the normalization of our kernel. The standard
archive contains eighteen six-dimensional records, and the single-class list
contains 186 rank-six representatives.

The order-four lattice $K^{+}_4$ agrees with the 155th rank-six representative
in the linked single-class list, with the same normalization. The rootless
order-eight lattice $K^{+}_{8,3}$ is isometric to \texttt{6QF.6.d}. Its determinant
is $512$, minimum is four, kissing number is thirty, and automorphism group
has order $384$. The other three order-eight kernels are not identified in
these records. As noted for order twenty-three, the entry \texttt{6QF.6.e} is
odd and does not match any of our even kernels.

For the other 119 modular classes, no matching entry was found in the records
checked. The comparison is restricted to explicit catalogue entries, including the
linked single-class data. Some lattices may already be known as direct sums
or rescalings of lower-rank lattices without a separate rank-six entry. Repeated kernels are counted
only once. In particular, the two descriptions in the $20C$ orbit share their thirteen
modular classes, whereas the two genera in the $15E$ and $24F$ orbits
contribute respectively nine and nineteen modular classes each.

\section{The order-twenty-three partition function from the charge lattice}
\label{app:order23-partition}

We derive~\eqref{eq:order23-partition} for the standard order-$23$ lift of $\Theta$ and an allowed shift
$v$ of exact order $23$ modulo $I^*$. The argument uses the self-orbifold property of the unshifted theory
and the fact that $23$ is prime. It parallels the elimination of the parent block and modular-orbit
rearrangement used for prime-order CHL strings in Appendix~A of~\cite{Bossard:2017wum}. Among the symmetries
in our classification, only the order-twenty-three case has prime order. We make no corresponding claim
for the composite-order examples.

\paragraph{The unshifted self-orbifold.}
Let $\mathcal C$ denote the parent heterotic $T^6$ CFT. The nontrivial eigenvalues of $\Theta$ on the
left-moving internal bosons are $e^{2\pi i j/23}$, $j=1,\ldots,22$. Their twisted vacuum has weight
\begin{equation}
 h_\Theta=\frac14\sum_{j=1}^{22}\frac{j}{23}\left(1-\frac{j}{23}\right)
 =\frac{22}{23},
\end{equation}
so the unshifted action satisfies level matching. To see its self-orbifold property, introduce an auxiliary
circle of covering radius $R$ and quotient by $\Theta$ combined with a translation by $2\pi R/23$.
In the first twisted sector, a state with circle momentum number $n=1$ and winding $w=1/23$ survives the
projection because the translation phase cancels the eigenvalue of the twisted vacuum. With
$p_{L,R}=n/R\pm wR/\alpha'$, at $R_*^2=23\alpha'$ it satisfies
\begin{equation}
 p_R=0,\qquad \frac{\alpha'p_L^2}{4}=\frac1{23},\qquad
 h_\Theta+\frac{\alpha'p_L^2}{4}=1.
\end{equation}
Together with its conjugate in the inverse twisted sector and the circle current, it generates
$SU(2)_{23}$. Indeed, for a canonically normalized chiral boson $X_L$, the currents have the form
$J^\pm\sim\sigma_\pm e^{\pm i\sqrt{2/23}X_L}$ and
$J^3=i\sqrt{23/2}\,\partial X_L$. The Weyl reflection changes the sign of the radius deformation
$J^3\bar J_{\mathrm{circle}}$ and gives the exact identification $R\leftrightarrow23\alpha'/R$.
The large-radius limit recovers $\mathcal C$, while the small-radius limit, after decompactifying the dual
circle, recovers $\mathcal C/\langle\Theta\rangle$. Hence these two CFTs are isomorphic.

Write $B^{(0)}_{a,b}$ for the torus block twisted by $\Theta^a$ with insertion $\Theta^b$, where
$a,b\in\mathbb F_{23}$. The self-orbifold identity implies
\begin{equation}\label{eq:order23-parent-elimination}
 B^{(0)}_{0,0}=\frac1{23}\sum_{a,b\in\mathbb F_{23}}B^{(0)}_{a,b},
 \qquad
 B^{(0)}_{0,0}=\frac1{22}\sum_{(a,b)\ne(0,0)}B^{(0)}_{a,b}.
\end{equation}
All block identities below are understood before setting the common supersymmetric factor to zero.

\paragraph{Eliminating the parent block.}
Let $B^{(v)}_{a,b}$ be the blocks for $g=\Theta\delta_v$, where
$\delta_v|Q\rangle=e^{2\pi i Q\cdot v}|Q\rangle$. Both orbifolds have the same uninserted parent block.
Substituting~\eqref{eq:order23-parent-elimination} gives
\begin{equation}\label{eq:order23-rearranged-sum}
 Z_v=\frac1{23}\sum_{(a,b)\ne(0,0)}
 \left(\frac1{22}B^{(0)}_{a,b}+B^{(v)}_{a,b}\right).
\end{equation}
For each $s=1,\ldots,22$, the fixed lattice of $\Theta^s$ is $I$ and its frame shape is $1^1 23^1$.
The untwisted inserted blocks are therefore
\begin{equation}
 \begin{split}
 B^{(0)}_{0,s}&=\mathcal U\,F(\tau)\Theta_I(\tau,\bar\tau),\\
 B^{(v)}_{0,s}&=\mathcal U\,F(\tau)
 \sum_{Q\in I}e^{2\pi i sQ\cdot v}q^{Q_L^2/2}\bar q^{Q_R^2/2},
 \qquad F(\tau)=\frac1{\eta(\tau)\eta(23\tau)}.
 \end{split}
\end{equation}
Consequently, the contribution of the line $(a,b)=(0,s)$ to~\eqref{eq:order23-rearranged-sum} is
\begin{equation}\label{eq:order23-seed}
 \begin{split}
 A_v&:=\frac1{23}\sum_{s=1}^{22}
 \left(\frac1{22}B^{(0)}_{0,s}+B^{(v)}_{0,s}\right)\\
 &=\mathcal U F(\tau)\sum_{Q\in I}
 \left(\frac1{23}\sum_{s=0}^{22}e^{2\pi i sQ\cdot v}\right)
 q^{Q_L^2/2}\bar q^{Q_R^2/2}
 =\mathcal U\,\frac{\Theta_{K_v}}{\eta(\tau)\eta(23\tau)}.
 \end{split}
\end{equation}
The character average restricts the sum precisely to $K_v=\ker_Iv=L_e^*$.

Since $23$ is prime, the nonzero pairs in $\mathbb F_{23}^2$ split into $24$ lines, each containing $22$
nonzero pairs. Modular transformations take the line $(0,s)$ to every other line. Thus
\eqref{eq:order23-rearranged-sum} is the sum of the $24$ modular images of $A_v$, represented by
$1,S,ST,\ldots,ST^{22}$, with the convention
$(A|\gamma)(\tau,\bar\tau)=A(\gamma\tau,\gamma\bar\tau)$.
This step requires prime order, since nontrivial powers at composite order can have different fixed lattices and
oscillator traces, and the same single-seed argument does not apply.

\paragraph{Poisson resummation and level matching.}
Using the signature $(0,6)$ and determinant $|\det K_v|=23^3$, Poisson resummation gives
\begin{equation}
 \begin{split}
 \Theta_{K_v}(-1/\tau,-1/\bar\tau)
 &=\frac{(i\bar\tau)^3}{23^{3/2}}\Theta_{L_e}(\tau,\bar\tau),\\
 F(-1/\tau)&=\frac{\sqrt{23}}{-i\tau}
 \frac1{\eta(\tau)\eta(\tau/23)}.
 \end{split}
\end{equation}
The common factor obeys
$\mathcal U(-1/\tau,-1/\bar\tau)=(-i\tau)(i\bar\tau)^{-3}\mathcal U(\tau,\bar\tau)$.
The modular weights cancel, leaving the numerical factor
$\sqrt{23}/23^{3/2}=1/23$. Hence
\begin{equation}
 (A_v|S)(\tau,\bar\tau)=\frac{\mathcal U}{23}
 \frac{\Theta_{L_e}(\tau,\bar\tau)}{\eta(\tau)\eta(\tau/23)}.
\end{equation}
Since $\mathcal U$ is invariant under $T$, adding the $23$ translates of this expression to $A_v$
proves~\eqref{eq:order23-partition}. The eta functions at $\tau+b$ retain their multipliers, whose phases
enter the level-matching projection.

Explicitly, define nonnegative integers $c_m$ by
\begin{equation}
 \frac1{\eta(\tau)\eta(\tau/23)}=\sum_{m\ge0}c_m q^{(m-1)/23}.
\end{equation}
The $T$-average selects $Q^2/2+(m-1)/23\in\mathbb Z$, and the result can equivalently be written as
\begin{equation}
 \mathcal Z_v=
 \frac{\Theta_{L_e^*}}{\eta(\tau)\eta(23\tau)}
 +\sum_{Q\in L_e}
 \sum_{\substack{m\ge0\\Q^2/2+(m-1)/23\in\mathbb Z}}
 c_m q^{Q_L^2/2+(m-1)/23}\bar q^{Q_R^2/2}.
\end{equation}
Here $23Q^2/2\in\mathbb Z$ follows from the level-$23$ discriminant form of $I$, $23v\in I$ and
$23v^2/2\in\mathbb Z$, so the finite average is indeed a projector. The first term and this projected
sum are a rearrangement of the full orbifold amplitude, not its original untwisted and twisted sectors
separately. All dependence on $v$ is through $L_e$ and its dual. Since these lattices are purely
right-moving, an integral isometry of the kernels preserves the theta functions as well. This proves the
claimed equality of one-loop partition functions for inequivalent shifts yielding isometric kernels,
without assuming that their operator products are equivalent.

\bibliographystyle{islands-unsrt}
\bibliography{Main4_refs}

\end{document}